%% file: sigconf.tex
\documentclass[sigconf]{acmart}
\pdfoutput=1
\usepackage{booktabs,balance} % For formal tables

\setcopyright{none}
\acmDOI{}

\acmISBN{}

\acmConference[]{April}{8}{2018}
\acmYear{}
\copyrightyear{}

\acmArticle{}
\acmPrice{}

\begin{document}
\title[All Reality
({\tt {\LARGE *}R})
]
{
All Reality: Virtual, Augmented, Mixed (X), Mediated (X,Y), and Multimediated Reality
%(AR, VR, MR, XR + iYR = {\tt {\Huge *}}R)
%is All Reality (``{\tt {\Huge *}}Reality'').\\
%\{VR, AR, MR, XR + iYR = ZR\} $\in$ {\tt {\Huge *}}R = "All R".
}
%\titlenote{Produces the permission block, and
%  copyright information}
%\subtitle{}
%\subtitlenote{}

\author{Steve Mann, Tom Furness, Yu Yuan, Jay Iorio, and Zixin Wang}
%\authornote{}
%\orcid{}
%\affiliation{%
%  \institution{}
%  \streetaddress{}
%  \city{}
%  \state{}
%  \postcode{}
%}
%\email{}

%\author{Yu Yuan}
%\authornote{}
%\affiliation{%
%  \institution{}
%  \streetaddress{}
%  \city{}
%  \state{}
%  \postcode{}
%}
%\email{}

%\author{Jay Iorio}
%\authornote{}
%\affiliation{%
%  \institution{}
%  \streetaddress{}
%  \city{}
%  \country{}
%}
%\email{}

\renewcommand{\shortauthors}{Mann, Furness, Yuan, Iorio, and Wang}

\begin{abstract}
The contributions of this paper are: (1) a taxonomy, framework, conceptualization, etc., of the ``Realities'' (Virtual, Augmented, Mixed, Mediated, etc.), and (2) some new kinds of ``reality'' that come from nature itself, i.e. that expand our notion beyond synthetic realities to include also phenomenological realities. 

VR (Virtual Reality) replaces the real world with a simulated experience (virtual world).  AR (Augmented Reality) allows a virtual world to be experienced while also experiencing the real world at the same time.  Mixed Reality 
%and ``eXtended'' Reality 
provides blends that {\em interpolate} between real and virtual worlds in various proportions, along a ``Virtuality'' axis, and {\em extrapolate} to an ``X-axis'' defined by Wyckoff's ``XR'' (eXtended reality), and Sony's X-Reality\texttrademark.
%especially in the context of extrapolated reality (e.g. Charles Wyckoff, as well as Sony's X-Reality\texttrademark \: and ``XR\texttrademark \:'' for video and mobile augmented reality.)

Mediated Reality goes a step further by mixing/blending {\em and also modifying} reality.  This modifying of reality introduces a second axis called ``Mediality''.  Mediated Reality is useful as a seeing aid (e.g. modifying reality to make it easier to understand), and for psychology experiments like Stratton's 1896 upside-down eyeglasses experiment.
%or seeing the world as a photographic negative.

We propose Multimediated Reality (``All Reality'' or ``All R'') as a multidimensional multisensory mediated reality that includes not just interactive multimedia-based ``reality'' for our five senses, but also includes additional senses (like sensory sonar, sensory radar, etc.), as well as our human actions/actuators.  These extra senses are mapped to our human senses using synthetic synesthesia.  This allows us to directly experience real (but otherwise invisible) phenomena, such as wave propagation and wave interference patterns, so that we can see radio waves and sound waves and how they interact with objects and each other.

%A special kind of lock-in amplifier was designed specifically for multimediated reality.  It can be used to scan a space using a 3D positioner and then the resulting wavefronts are experienced on a wearable computing system, as well as on video displays in the environment. Wearers of the computer system can experience a multidimensional space (X, Y, Z, time, phase, real, imaginary, etc.). Those not wearing the computer system can still experience a reduced dimensional slice through this multidimensional space using ambient display media.

Multimediated reality is multidimensional, multimodal, multisensory, and multiscale, including not just ``wearables'' but also smart buildings, smart cities, etc..  It is also multidisciplinary, in that we must consider not just the user, but also how the technology affects others, e.g. how its physical appearance affects social situations.  Finally, it is multiveillant (surveillance, data-veillance, and other ``veillances'').  For example, cameras in multimediated reality devices affect the privacy of non-users of the technology as well.

%\footnote{This is an abstract footnote}
\end{abstract}

%\begin{CCSXML}
%<ccs2012>
%<concept>
%<concept_id>10002944.10011123.10011131</concept_id>
%<concept_desc>General and reference~Experimentation</concept_desc>
%<concept_significance>100</concept_significance>
%</concept>
%</ccs2012>
%\end{CCSXML}

%\ccsdesc[100]{General and reference~Experimentation}

\keywords{Virtual reality, VR, augmented reality, AR, mixed reality, mediated reality, wave propagation, education, physics, lock-in amplifier, standing waves, sitting waves}

\maketitle

\input{body-conf}

\balance

\bibliographystyle{ACM-Reference-Format}


\newcommand{\noopsort}[1]{}

\end{document}

%% file: body-conf.tex
\section{Historical Background \& Context}
\subsection{Virtual, Augmented, Mixed, and X-Reality}%\texttrademark}
VR (Virtual Reality) is a computer-generated simulation of a realistic experience.
Typically VR blocks out the real world (``Reality'') and replaces it with a ``Virtual'' world.  The virtual world may be generated by a computer, or by interactively playing back recorded media.  An example of the latter is the Aspen Movie Map of 1978 that used computers to play back analog laser disk recordings to render an interactive virtual world as hypermedia~\cite{mohl1981cognitive}, or, more recently, Google Street View with Earth VR.

AR (Augmented Reality) is a similar concept, but instead of
blocking out reality, the computer-generated content is
added onto, or embedded into, the real world experience, so that both can be
experienced together~\cite{azuma1997survey}.

\begin{figure}
\includegraphics[width=\columnwidth]{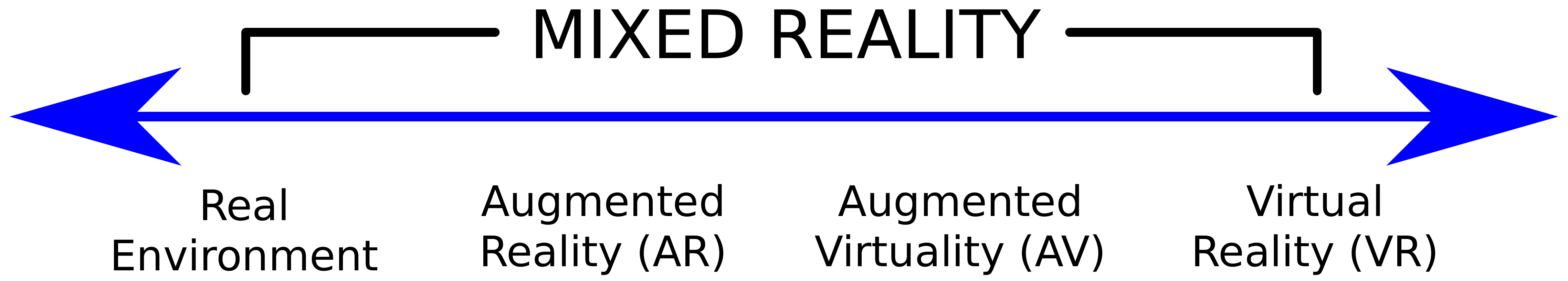}
\vspace{-.2in}
\caption{
Above: Mixed Reality Continuum, adapted from Milgram and Kishino, 1994~\cite{milgram94}. The blue arrow is suggestive of a one-dimensional ``slider'' or ``fader'' that ``mixes'' between the Real world and the Virtual world, as illustrated below:
        }
\includegraphics[width=\columnwidth]{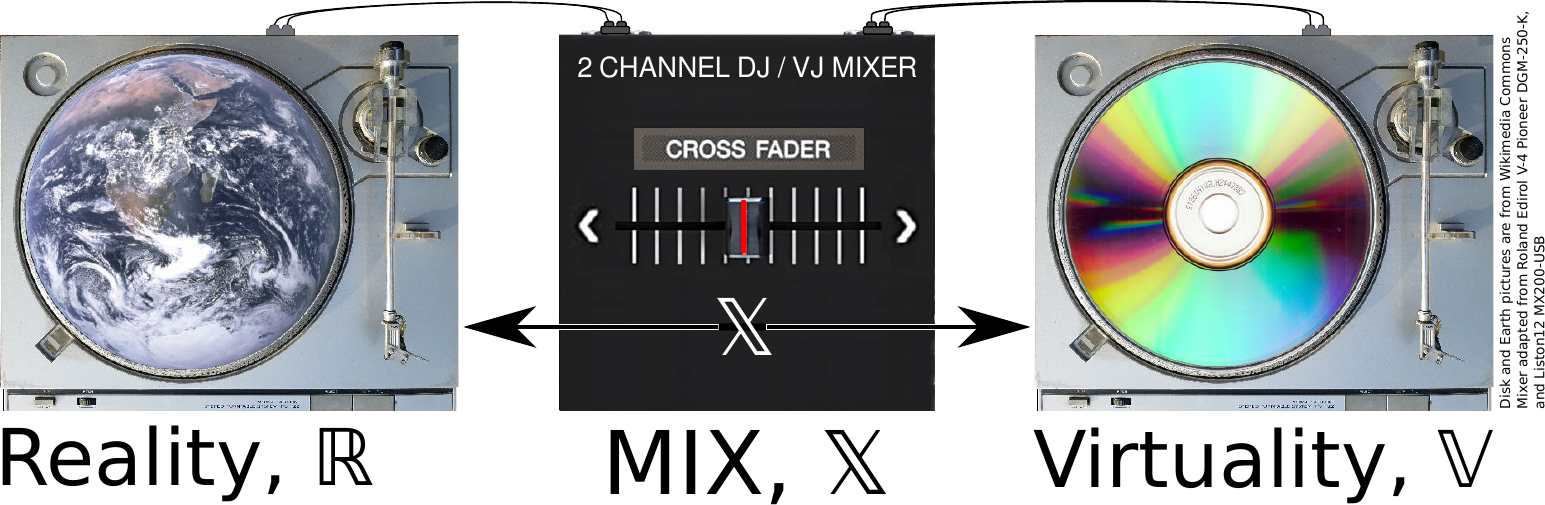}
\vspace{-.25in}
\caption{Disk Jockey (DJ) Mixer Metaphor:  Imagine two record players (turntables), feeding into an audio/video mixer.
Real-world and virtual world mixtures are selected by sliding a
one-dimensional ``fader'' left or right.
%, eXtended, eXpanded, or virtual reality
This allows us to choose various points along an "$\mathbb{X}$" axis between
the extremes of Reality, "$\mathbb{R}$", and Virtuality, "$\mathbb{V}$".
}
\label{fig:milgram}
\end{figure}

It has been suggested, by Milgram and Kishino~\cite{milgram94}, that Augmented Reality exists along a continuum between the real and virtual worlds, giving rise to ``mixed reality''
(see Fig.~\ref{fig:milgram}).
In this context we can think of AR as a setting on a ``mixer'' or ``fader'' or ``slider'' that is somewhere between reality and virtuality.

This ``slider'' is analogous to the ``X-axis'' of an X-Y plot or graph,
treating ``X'' as a mathematical variable that can assume any quantity
on the real number line.  Thus mixed reality is sometimes referred to as ``X-reality'' or ``XR''~\cite{intelligentimageprocessing,paradiso2009guest,coleman2009using}.
Specifically, a 2009 special issue of IEEE ``PERVASIVE computing''
on ``Cross-Reality Environments''
defines X-reality as a proper subset of Mixed Reality.
Paradisio and Landay define ``cross-reality'' as:
\begin{quote}
``the union between ubiquitous sensor/actuator networks and
shared online virtual worlds ....
We call the ubiquitous mixed reality environment that comes from
the fusion of these two technologies cross-reality.''~\cite{paradiso2009guest}
\end{quote}
In that same issue of IEEE ``PERVASIVE computing'',
Coleman defines ``cross-reality'' and
``X-Reality'' as being identical:
\begin{quote}
``Cross-reality (also known as x-reality) is an informational or
media exchange between real-and virtual-world systems.''~\cite{coleman2009using}.
\end{quote}

XR as extrapolation (``extended reality'' or ``extended response'') dates back as early as 1961 when Charles Wyckoff filed a patent for his ``XR'' film which allowed people to see nuclear explosions and other phenomena beyond the range of normal human vision~\cite{wyckoff1962experimental,wyckofftr, life1966}.  In 1991, Mann and Wyckoff worked together to build ``XR vision'' devices into wearable computers (AR/VR headsets, etc.) for human augmentation and sensory extension by way of High Dynamic Range (HDR) imaging blended with virtual/augmented reality~\cite{intelligentimageprocessing}.

The terms ``XR'', ``X-Reality, ``X-REALITY, and ``XREALITY appear as trademarks registered to Sony Corporation, filed in 2010, and used extensively in the context of mobile augmented reality across Sony's ``Xperia'' X-Reality\texttrademark \: for mobile products, as shown in Fig.~\ref{fig:sony} below:\\
\begin{figure}[h]
\vspace{-.3in}
\begin{center}
\includegraphics[width=.85\columnwidth]{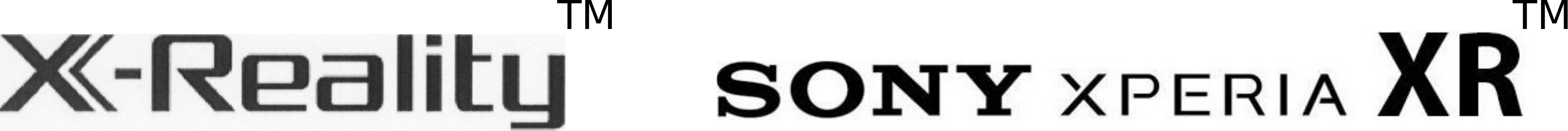}
\end{center}
\vspace{-.16in}
\caption{Sony's trademarked X-Reality and XR
\vspace{-.2in}
}
\label{fig:sony}
\end{figure}

\noindent
Sony's use of XR and X-Reality is consistent with the Wyckoff-Mann conceptualization of extended human sensory perception through high dynamic range.

\begin{figure}
\includegraphics[width=.75\columnwidth]{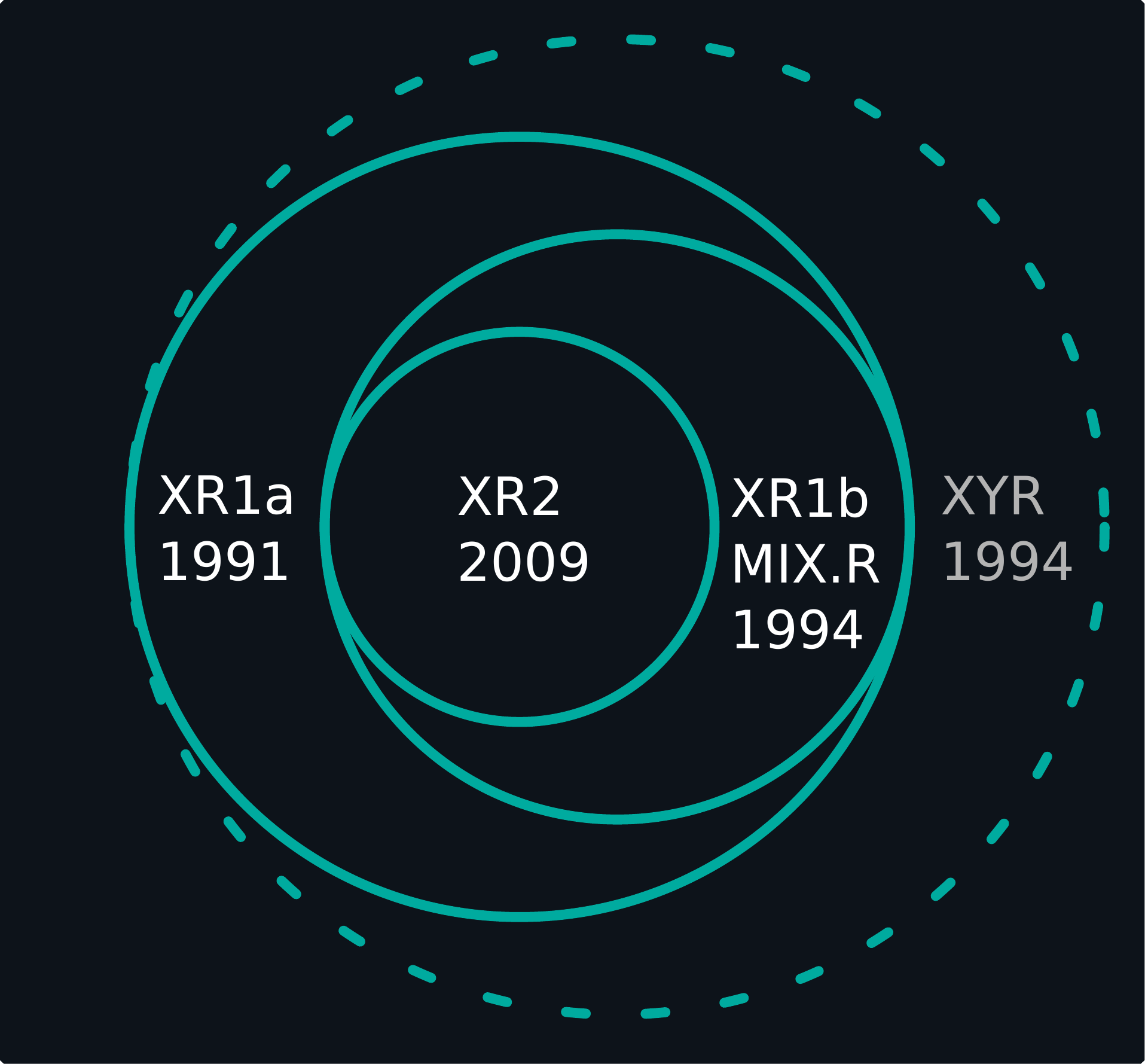}
\caption{Taxonomy (Venn diagram) of the three definitions of XR (X-Reality), as well as XY-Reality which will be introduced in the next section.  Leftmost: XR1a introduced in 1991 is the most general of the X-realities.  XR1b is identical to mixed reality.  XR2, introduced in 2009, is a subset of mixed reality, specifically the combination of ``wearables'' and ``smart cities''.}.
\label{fig:xr}
\end{figure}

There is some confusion, though, since XR (X-Reality) now has at least three definitions, one in which it is a proper {\em superset} of mixed reality, another in which it {\em is} mixed reality, and another in which it is a proper {\em subset} of mixed reality.
We shall enumerate and classify these, chronologically, as follows:
\begin{itemize}
\item {\bf Type 1 XR/X-Reality} in which ``X'' as a mathematical variable, i.e. any number on the real number line, that defines an axis for either:
\begin{itemize}
\item {\bf Type 1a XR/X-Reality: extrapolation}, i.e. XR/X-Reality in Wyckoff-Mann sense (1991), as technologies that extend/augment/expand human sensory capabilities through wearable computing.  In this sense ``X'' defines an axis that reaches past ``reality''.
\item {\bf Type 1b XR/X-Reality: interpolation}, i.e. XR/X-Reality in the Milgram sense (1994), as technologies that augment human senses by creating a blend (mixture) between the extremes of reality and virtuality.  In this sense ``X'' defines an axis that miXes (interpolations) between reality and virtuality.
\end{itemize}
\item {\bf Type 2 XR/X-Reality} in which ``X'' means ``Cross'' in the Paradiso-Landay/Coleman sense (2009),
i.e. as a form of mixed reality (a proper subset of mixed reality)
in which the reality portion comes from sensor/actuator networks, and the virtuality portion comes from shared online virtual worlds.
\end{itemize}
The taxonomy of these three definitions of XR/X-Reality is summarized as a Venn diagram in Fig.~\ref{fig:xr}, showing also XY-Reality (XYR) which will be defined in the next section.

What these definitions of XR/X-Reality all have in common is that XR/X-Reality defines an ``X-axis'' defining a number line that passes through both ``reality'' and ``virtuality''.

\subsection{Mediated Reality (X-Y Reality)}
Many technologies function as an intermediary between us and the environment around us.  Technology can modify or change (mediate) our ``reality'', either as a result of deliberate design of the technology to mediate reality, or sometimes as an accidental or unintended side-effect.
These two variants of mediated reality are further discussed below.

\subsection{Deliberately mediated reality}
Examples of deliberate modification of reality include the upside-down eyeglass invented 122 years ago by George Stratton to study the effects of optically mediated vision on the brain~\cite{stratton96}.  More recently others have done similar experiments with deliberate mediation of reality, such as left-right image reversing eyeglasses~\cite{dolezal82}.
Multimedia devices such as hand-held camera viewfinders have also been used to study long-term adaptation to a photographic negative view of the world in which light areas of the image are made dark, and dark areas of the image are made light~\cite{anstis92}.  Computer-mediated reality has also been explored~\cite{mann260, presenceconnect}.  

Mediated Reality is not just for psychology experiments, though.  It has many practical everyday applications such as eyeglasses that filter out advertisements, and, more generally, helping people see better by getting rid of visual clutter.  HDR (High Dynamic Range) welding helmets use computer vision to {\em diminish} the otherwise overwhelming brightness of an electric arc, while {\em augmenting} dark shadow detail.  In addition to this Mediated Reality the HDR welding helmet also adds in some virtual content as well~\cite{davies2012quantigraphic}.

Mediated Reality has also been examined in the context of wearable computing, prosthesis, and veillance (surveillance, sousveillance, metaveillance, and dataveillance)~\cite{tan13, ganascia2010generalized}.

\subsection{Unintentionally Mediated Reality}
In Augmented Reality there is often an attempt made to not alter reality at all.
But when we experience augmented reality through a smartphone or tablet, or by using video see-through eyeglasses, the simple fact that we have a technology between us and our outside world means that the virtual objects overlaid onto reality are actually being overlaid onto an unintentionally modified reality (ie. both the virtual and real objects are presented by a video display device).
Thus the Mediated Reality framework is directly applicable to the research and understanding of video-see-through implementations of AR.

It has also been observed that head-mounted displays
can cause semi-permanent or permanent and lasting harm (e.g. brain damage) or good, such as in the treatment of 
PTSD (Post Traumatic Stress Disorder), phobias, and in treating (reversing the effects of) brain damage~\cite{wiederhold2005virtual,stanney1997psychometrics,rothbaum2010virtual,rizzo2015virtual,rizzo1997virtual}.
The fact that HMDs can both damage the brain as well as treat brain damage suggests that we need to be extra careful when using technology as an intermediary, and that there is a special importance to mediality in general.

\subsection{The Mediated Reality (X,Y) Continuum}
Summarizing the previous two subsections: consider either of the following:
\begin{itemize}
 \item devices and systems designed to intentionally modify reality;
 \item the unintended modification of reality that occurs whenever we place any technology between us and our surroundings (e.g. video-see-through Augmented Reality).
 \end{itemize}
Both of these situations call for at least one other axis beyond the mix between reality and virtuality.

\begin{figure}
\includegraphics[width=\columnwidth]{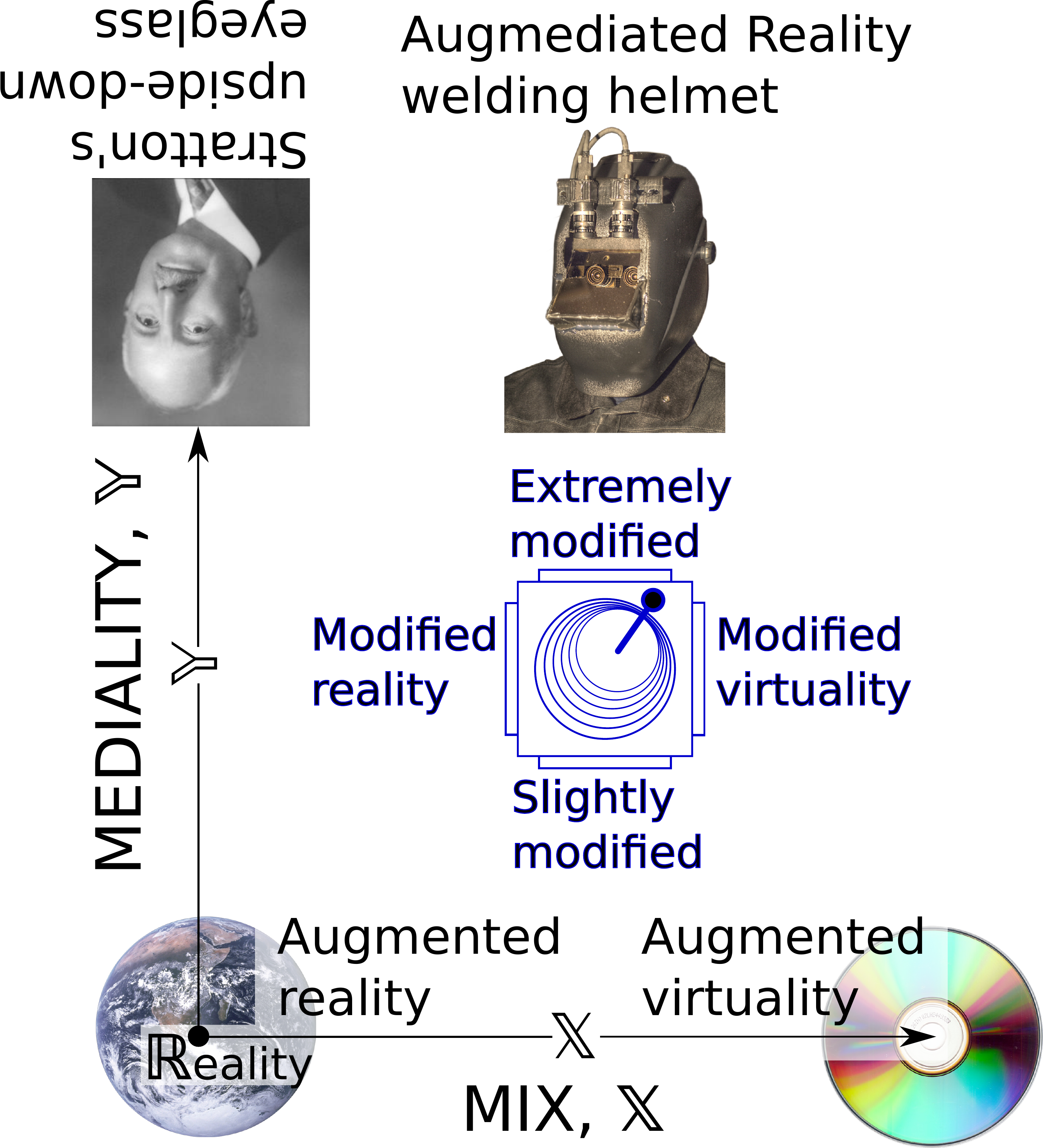}
\caption{Mediated Reality (X,Y) Continuum:
There exists a continuum in both the degree to which reality can be virtualized, as well as the degree to which it can be modified.
The ``MIX'' axis (``$\mathbb{X}$'' axis) runs left-to-right (reality to virtuality).
The ``MEDIALITY'' axis (`$\mathbb{Y}$'') runs bottom to top (Slightly modified to Extremely modified).
George Stratton's upside-down glass is an example of a wearable eyeglass technology that involves no virtuality but a great deal of mediality, and thus occupies an area in the upper left.
The EyeTap HDR welding helmet~\cite{davies2012quantigraphic} an example of extreme reality modification (mediality), that also involves a moderate amount of virtuality.
The amount of virtuality it has is about the same as a typical augmented reality setup, so it exists near the top middle of the space.
This top middle area of the Continuum is sometimes called ``Augmediated Reality'' (Augmented Mediated Reality)~\cite{minsky2013society, janzen2014walking}.
}
\label{fig:mediality}
\end{figure}

Moreover the above are just two of the many more examples of ``reality'' technologies that do not fit into the one-dimensional ``mixer'' of Fig.~\ref{fig:milgram}, and thus we need at least one additional axis when describing technology that specifically modifies reality.
For this reason, Mediated Reality~\cite{mann260, tan13, haller2005loose, tang2002seeing} has been proposed.
See Fig~\ref{fig:mediality}.
In this Mediated Reality taxonomy (continuum), there are two axes: the virtuality axis (``X'') exactly as proposed by Milgram, and a second axis, the Mediality axis (``Y'').  This allows us to consider other possibilities like mediated-augmented-reality (``augmediated reality'')~\cite{davies2012quantigraphic} (e.g. HDR welding helmets), as well as
mediated virtuality (e.g. taking an existing VR system and then flipping the image upside-down, to allow us to repeat George Stratton's 1896 upside-down eyeglass experiment but in a virtual world).

\section{Multimediated Reality}
\subsection{Technologies for sensory attenuation}
The Milgram Continuum (Fig~\ref{fig:milgram}) (Milgram 1994~\cite{milgram94})
and the Mann Continuum (Fig~\ref{fig:mediality}) (Mann 1994~\cite{mann260})
both place reality at the left or the lower left, i.e. the``origin'' in Cartesian coordinates.

Neither Milgram's nor Mann's Continuum directly addresses visual sensory attenuation technologies like sunglasses and sleep masks, or attenuation of other senses by such technologies as ear plugs or sensory attenuation tanks
(also known as ``sensory deprivation tanks'' or ``flotation tanks'').

Other visual useful sensory attenuation devices include the sun visor of a car, the brim of a baseball cap, or the ``blinders'' attached to a horse's bridle so that the horse is not distracted by peripheral motion cues.
%Even without such technology, people sometimes ``cup'' their hands
%around their eyes to briefly restrict (diminish) their peripheral
%vision in order to focus their attention on a specific object of
%interest.

\begin{figure}
\includegraphics[width=\columnwidth]{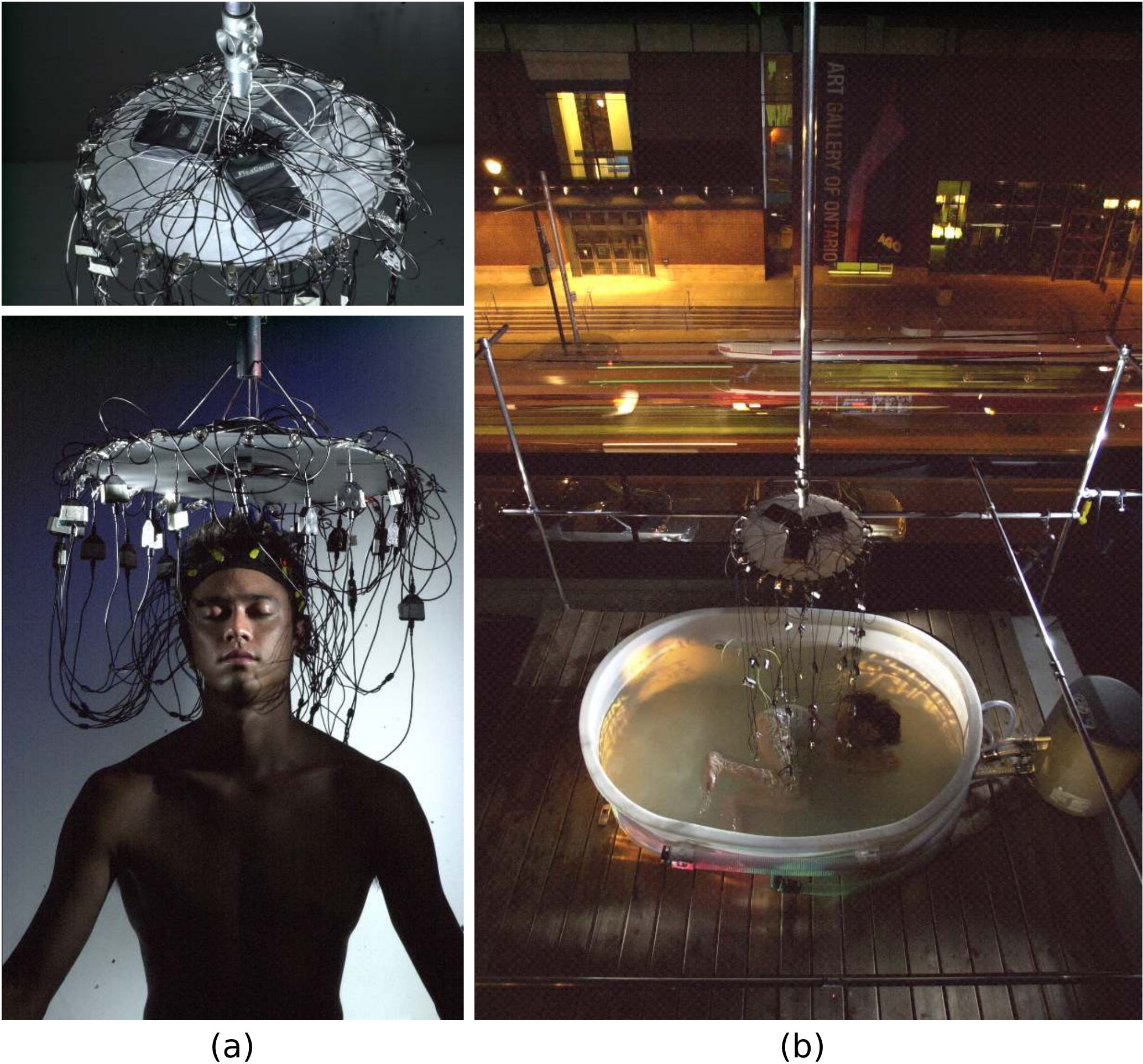}
\vspace{-.3in}
\caption{Submersive Reality (SR): Meditation in a VR/AR/MR Flotation tank. (a) Waterproof 24-electrode underwater EEG (ElectroEncephaloGram) system is part of the many multidimensional inputs to a real-time computer system (b) Interactive multimedia VR/AR/MR environment with water pumps, waves, vibration, lighting, and data projectors controlled by brainwave mediation/meditation\protect\cite{teletubs,icmc2007performance}.}
\includegraphics[width=\columnwidth]{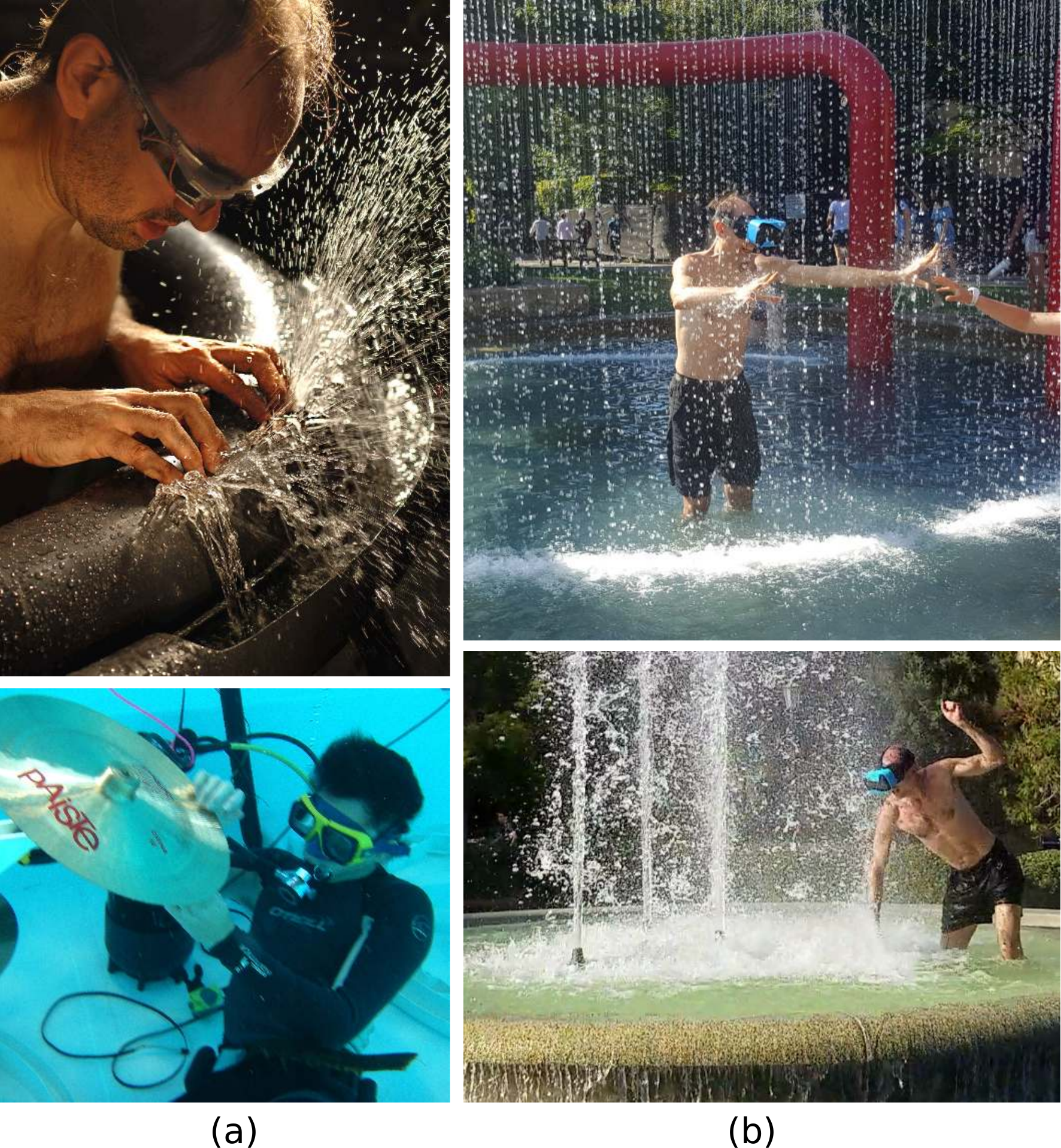}
\vspace{-.3in}
\caption{We also developed submersive reality for (a) musical performances, underwater concerts, etc., and (b) physical fitness, e.g. swimming in virtual environments and interaction with water jets~\protect\cite{teletubs,icmc2007performance}.}
\label{fig:tubs}
\end{figure}

Sensory attenuation technologies form an underexplored yet richly interesting space for technology.  Consider for example, some of the following possibilities:
\begin{itemize}
\item Interactive sleep masks for shared lucid dreaming.
\item Interactive multimedia bathing environments like computer-mediated sensory attenuation tanks, and interactive relaxation tanks that use water
for sensory attenuation (Mann 2004), as well as ``immersive multimedia'' and ``Fluid User Interfaces''~\cite{mann2005fl} that use water to alter the senses in conjunction with interactive multimedia.  See Fig.~\protect\ref{fig:tubs}.
\item Interactive darkroom experiences such as interactive lightpainting with reality-based media such as persistence-of-exposure and ``Phenomenological Augmented Reality'' (e.g. being able to see radio waves and see sound waves by way of a darkroom environment with eyes adjusted to the dark).
\end{itemize}
%Matt Kim might be interested in sensory deprivation

\subsection{Multimedia in Photographic Darkrooms}
A light bulb waved around in a dark room will tend to create the visual appearance or impression of shapes, curves, patterns, etc., by way of a ``persistence-of-exposure'' effect in human vision as well as in photographic or videographic media.  There is a long history of photographic ``lightpainting''
(http://lpwa.pro/event/15).
There is also a well established ``flow arts'' community doing artistic dance in a dark environment with light sources, e.g. LED (Light Emitting Diodes), as well as ``fire spinning'' and juggling light-emitting objects as a medium of creative expression.  Flow art is similar to lightpainting but for direct viewing rather than through photography.  Some tools (specialized light sources) and devices are used for both lightpainting and flow arts.

The tradition of darkness 
(sensory attenuation) combined with sensory media (e.g. controlled lighting) dates back to the early days of theatre.  Theatrical productions typically take place in a space in which all or most of the walls are painted black, and there is usually a black floor, and black curtains, such that lighting can be controlled carefully.  In fact the world's first use of the term ``Virtual Reality'' came from theatre in 1938~\cite{artaud}.

The tradition of sensory attenuation and controlled sensory stimulus connects well to multimedia:
Morton Heilig produced the ``Sensorama'' (U.S. Pat. \#3050870), a multi-sensory experience which was also the world's first ``3D film''.
%It also included an oscillating fan so the participant 
%would feel actual wind blowing.

\subsection{Multimediated Reality Darkroom}
In the 1970s, the idea of an interactive darkroom was taken a step further, by conducting a series of experiments to make otherwise invisible phenomena visible.
These experiments involved light bulbs connected to the output of powerful amplifiers that were
driven by transducers or antennae that sensed a physical quantity of interest.

In one example, a light bulb was used to ``sweep'' for video ``bugs'' and the light bulb glowed more brightly when in the field of view of a surveillance camera, than it did when not in the camera's field of view (Mann 2014).
This works very simply by using video feedback: a receive antenna is connected to the input of a very sensitive lock-in amplifier with extremely high gain.  The lock-in amplifier was specially designed to drive high capacity loads such as powerful electric light bulbs (1500 to 2500 watts). 

\begin{figure}
\includegraphics[width=\columnwidth]{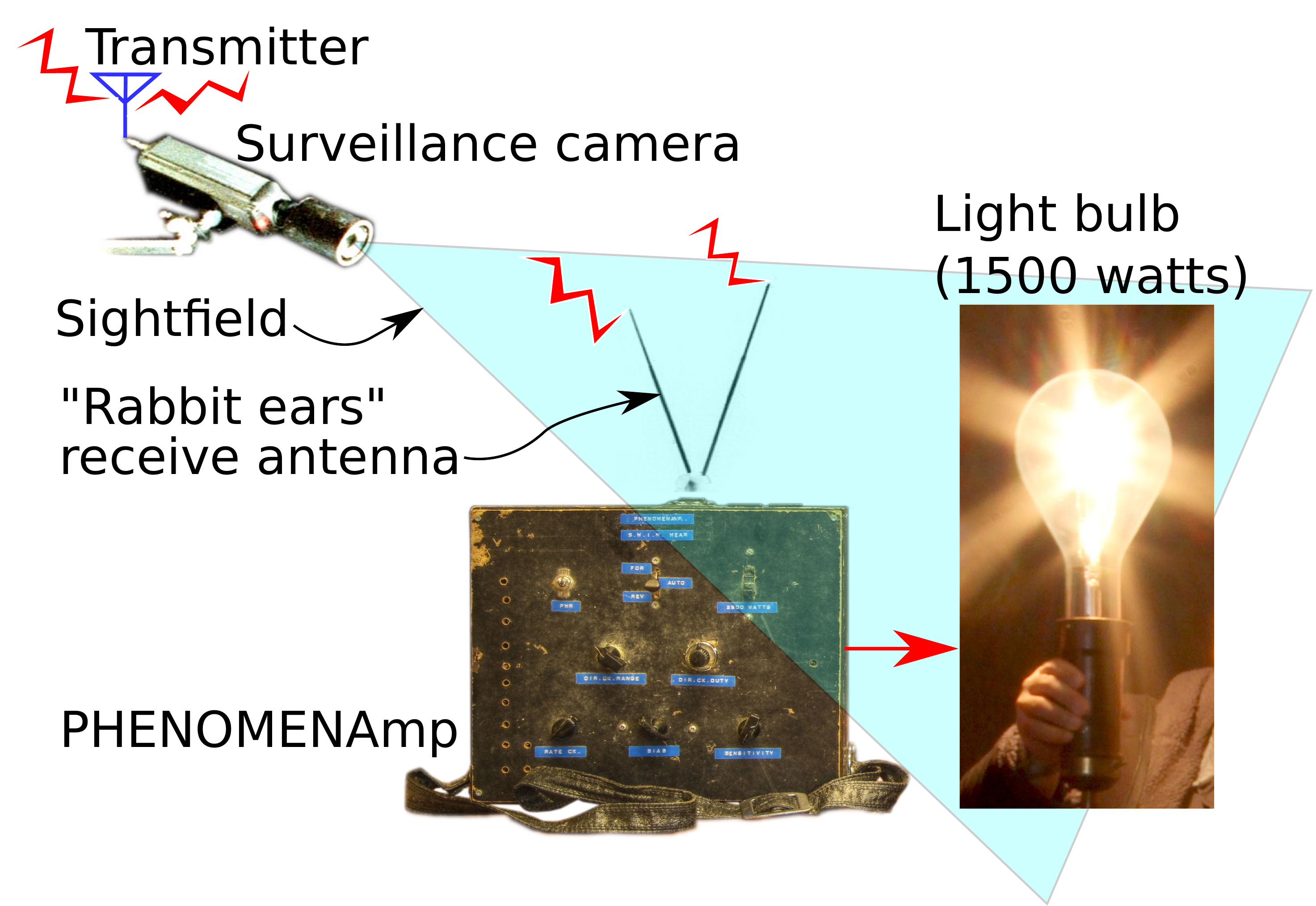}
\includegraphics[width=\columnwidth]{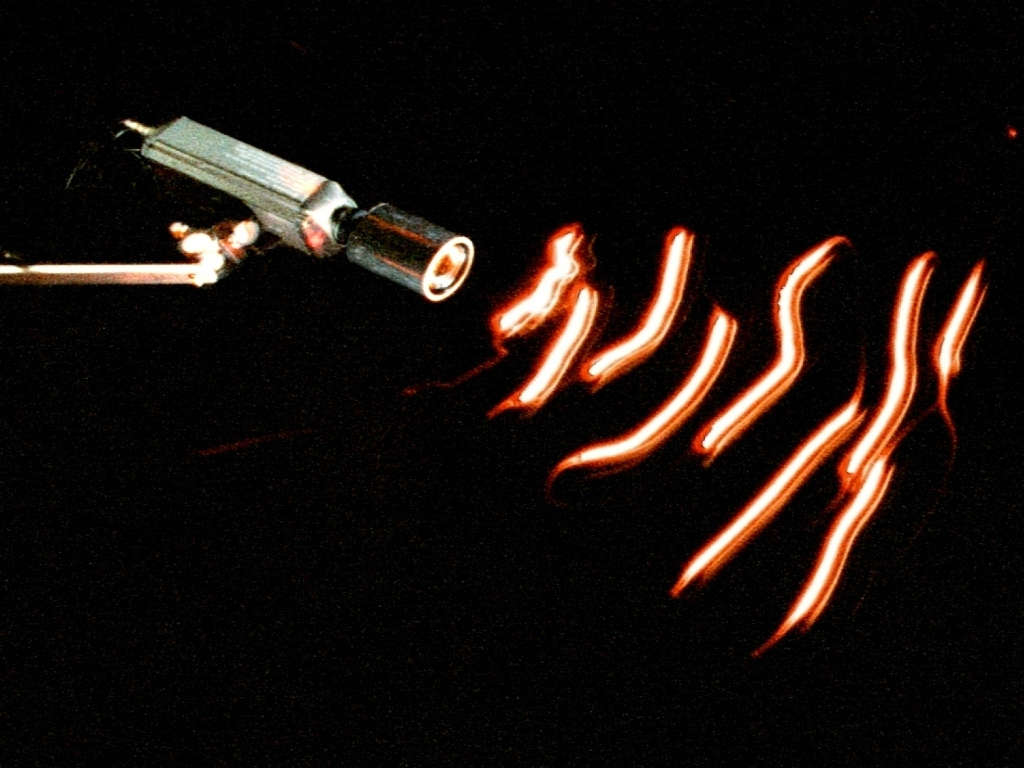}
\caption{A very sensitive lock-in amplifier picks up weak radio signals and drives a powerful light bulb causing video feedback.  The amplifier ``locks in'' on the signal that is due to itself, and thus glows brilliantly whenever it is within the field of view of the surveillance camera.  In this way, waving the light bulb back and forth ``paints'' out an image that reveals the otherwise hidden ``sightfield'' of the camera.}
\label{fig:metaveillance}
\end{figure}

When the light shines on a camera it causes the camera to exhibit small but measurable changes, causing video feedback.  Waving the light bulb back and forth in front of the camera makes the ``sightfield'' of the camera visible.  See Fig.~\ref{fig:metaveillance}.
It has been suggested that this is a form of augmented reality (Mann 2014) but it is a special kind of reality in the sense that it comes directly from nature itself.  Unlike many other forms of reality, it does not come from a computer simulation.  
Here the light bulb filament has a dual role: it is both the mechanism by which a physical quantity is sensed, and it is also the display mechanism.  Therefore, due to the fundamental physics of the situation, the alignment between the ``real'' physical world, and the ``augmented'' world is exact (there is no need for any tracking mechanism since the process itself is self-tracking).

We proffer to call this Phenomenological Reality, because it makes visible true physical quantities by way of directly physical means, i.e. a direct connection between a physical quantity and the sensed quantity.

\begin{figure}
\includegraphics[width=\columnwidth]{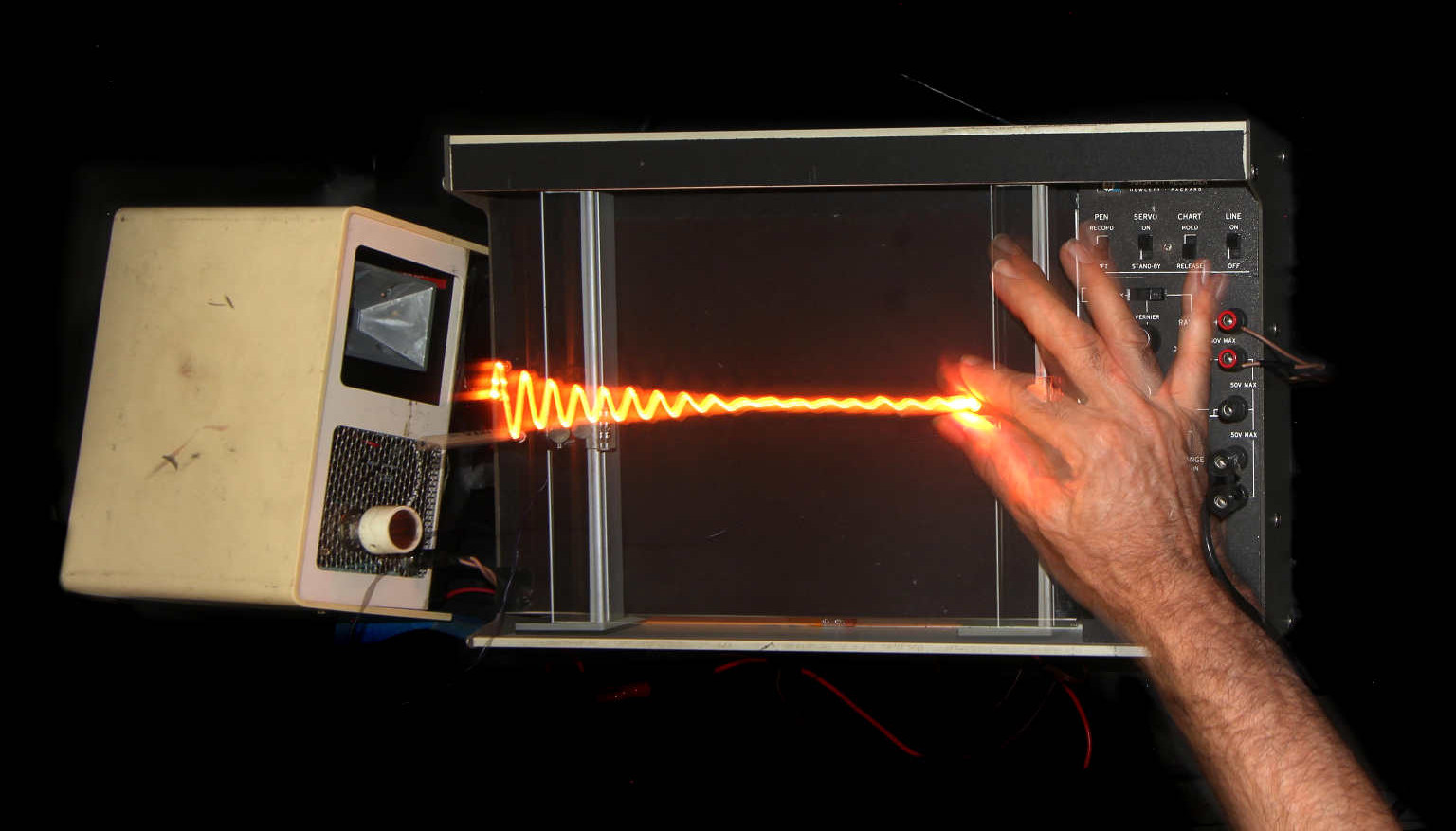}
\caption{The electrical signal from a microwave motion sensor (radar) is received by an antenna that moves together with a light bulb on an XY plotter.  The bulb traces out the waveform of the received signal in exact alignment with physical reality.  For example, we can measure the distance between cycles of the waveform as being exactly (300,000,000 m/s) / (10.525 GHz) = 2.85cm.  Note the hand grasping the bulb, making it possible for a human user to feel the radio wave in addition to seeing it.  Thus we have a visual and haptic multimediated reality experience that extends our senses beyond the five human senses, and maps this ``sixth sense'' (radio) onto two or three (we can also hear it) existing senses.}
\label{fig:plotter}
\end{figure}

\begin{figure*}
\includegraphics[height=2in]{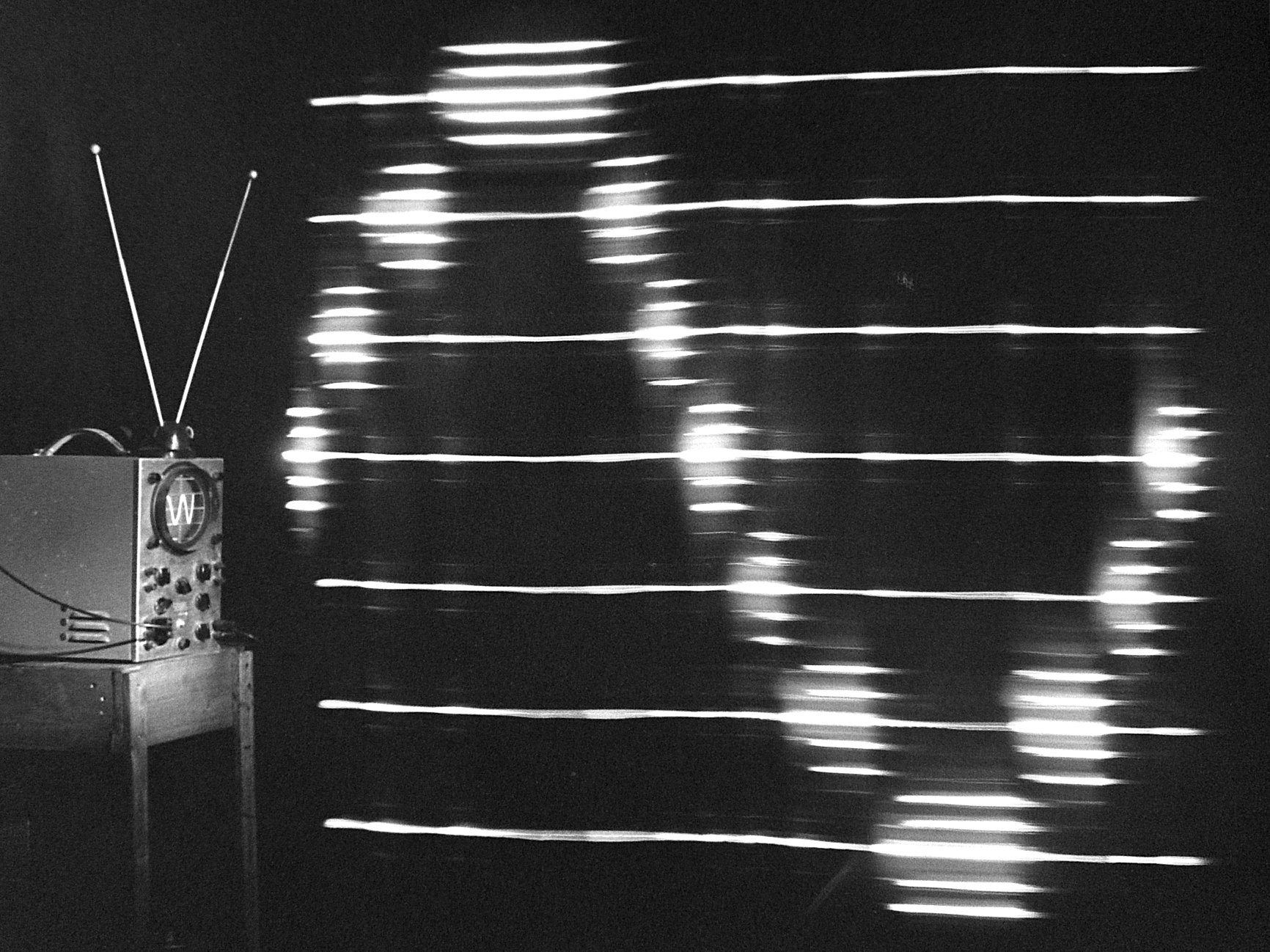}
\includegraphics[height=2in]{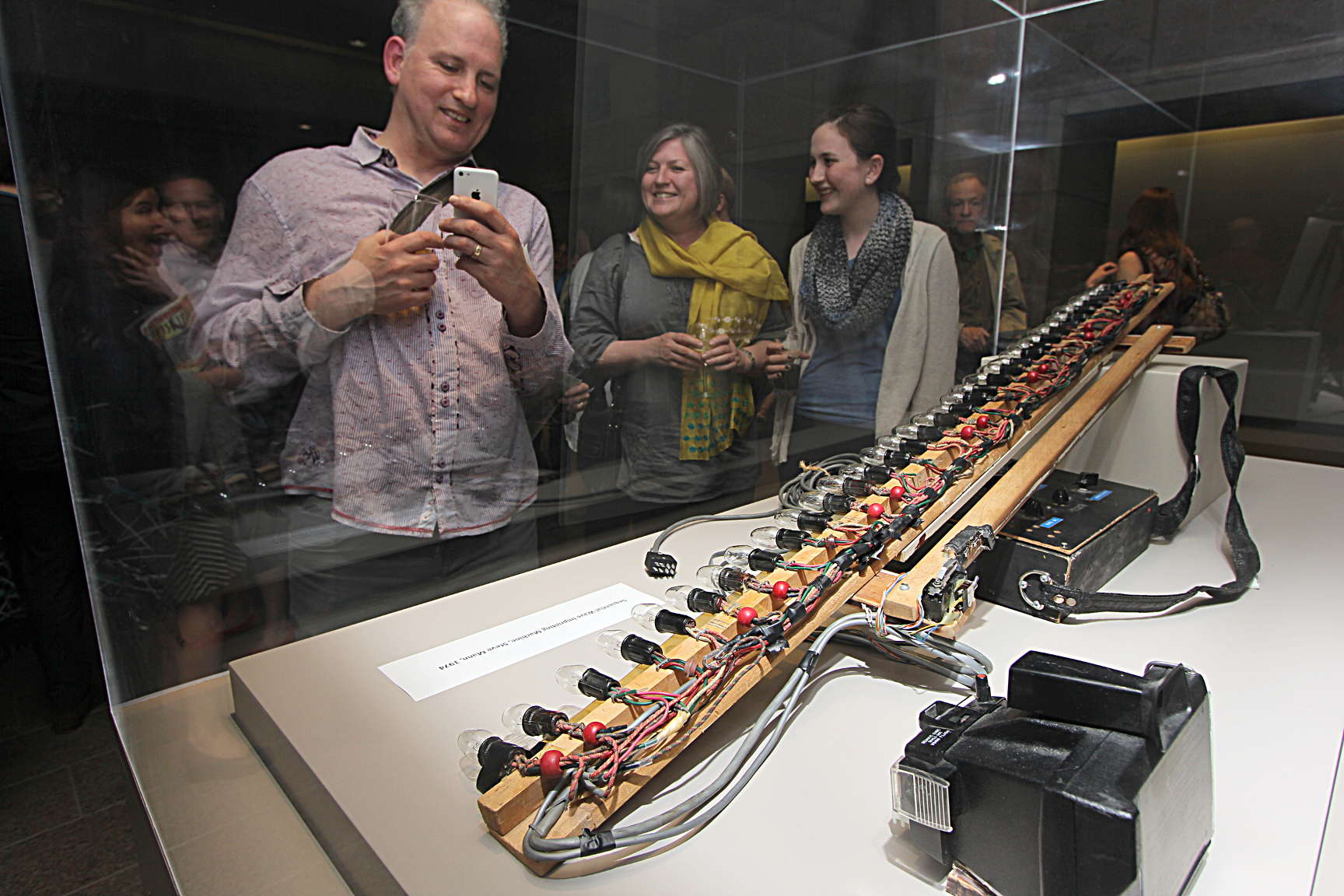}
\includegraphics[height=2in]{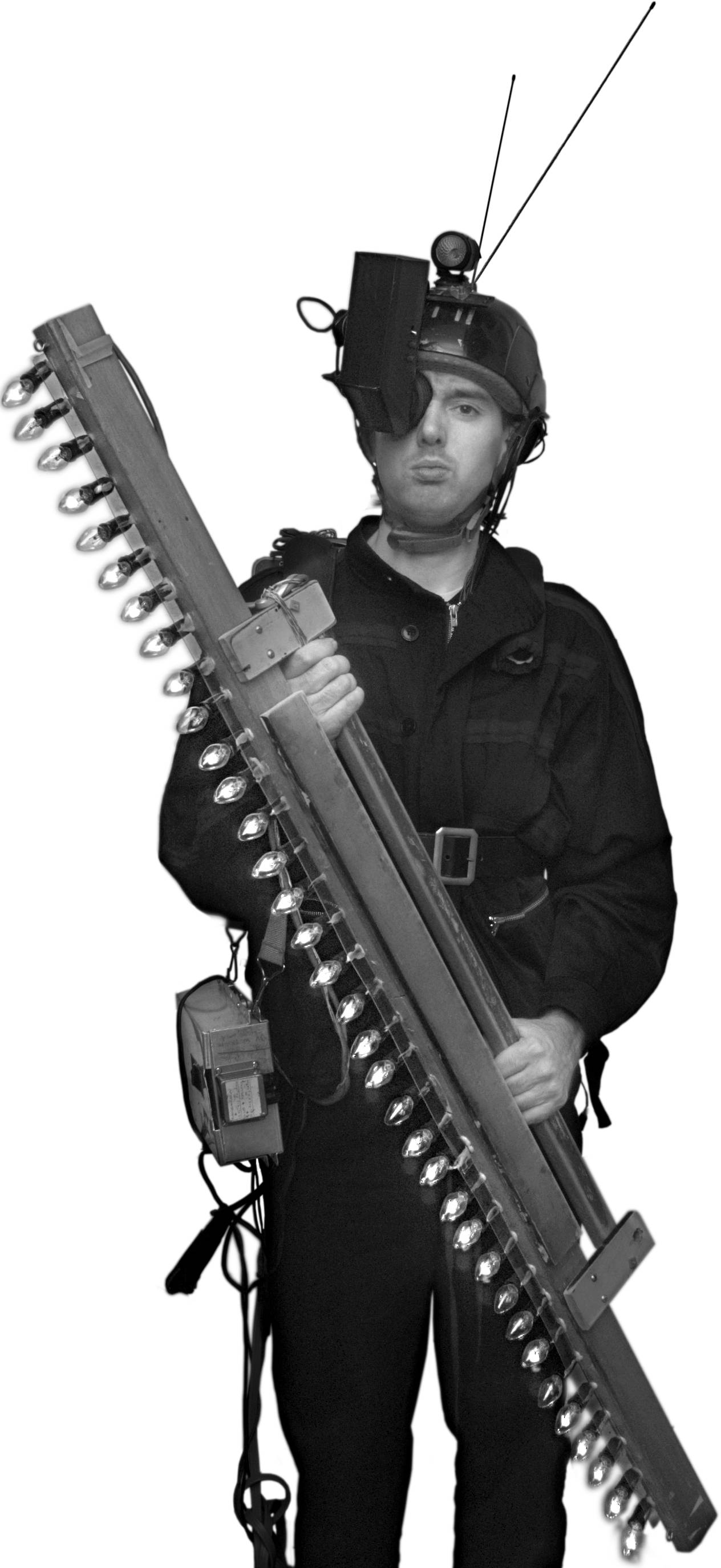}
\caption{Early photograph of electromagnetic radio waves using the Sequential Wave Imprinting Machine (S. Mann, July 6, 1974, from Wikimedia Commons).  Center: Sequential Wave Imprinting Machine exhibited at Smithsonian Institute and National Gallery, showing the linear array of 35 electric lights and the early wearable computer with lock-in amplifier and antenna sensor.  Right: Early (1980) version of the apparatus with early wearable multimedia computer prototype.
}
\label{fig:swim}
\end{figure*}

When the quantity we wish to sense is not the same thing as light itself, we can use a separate sensor but mount it directly to the light bulb.  For example, Fig.~\ref{fig:plotter}
shows a light bulb being moved by an XY plotter driven by the output of a lock-in amplifier.  It is important to emphasize here that the path the light bulb takes is in perfect alignment with the physical quantity, and thus provides us with a direct glimpse into the physical nature of the signal that is being shown.
Here the quantity that we're measuring is electromagnetic energy, phase-coherently, with an antenna affixed directly to the light bulb.  A special technique is used to transform the space in such coordinates that the speed of light is exactly zero, and thus we can see the radio wave sitting still.
The first such photograph of radio waves was made using a linear array of light bulbs, controlled by a wearable computer, in 1974~\cite{mann2015par}. See Fig\ref{fig:swim} (See~\cite{kineveillance} for an explanation of the methodology of photography of radio waves and sound waves.)

\begin{figure*}
\includegraphics[height=2.5in]{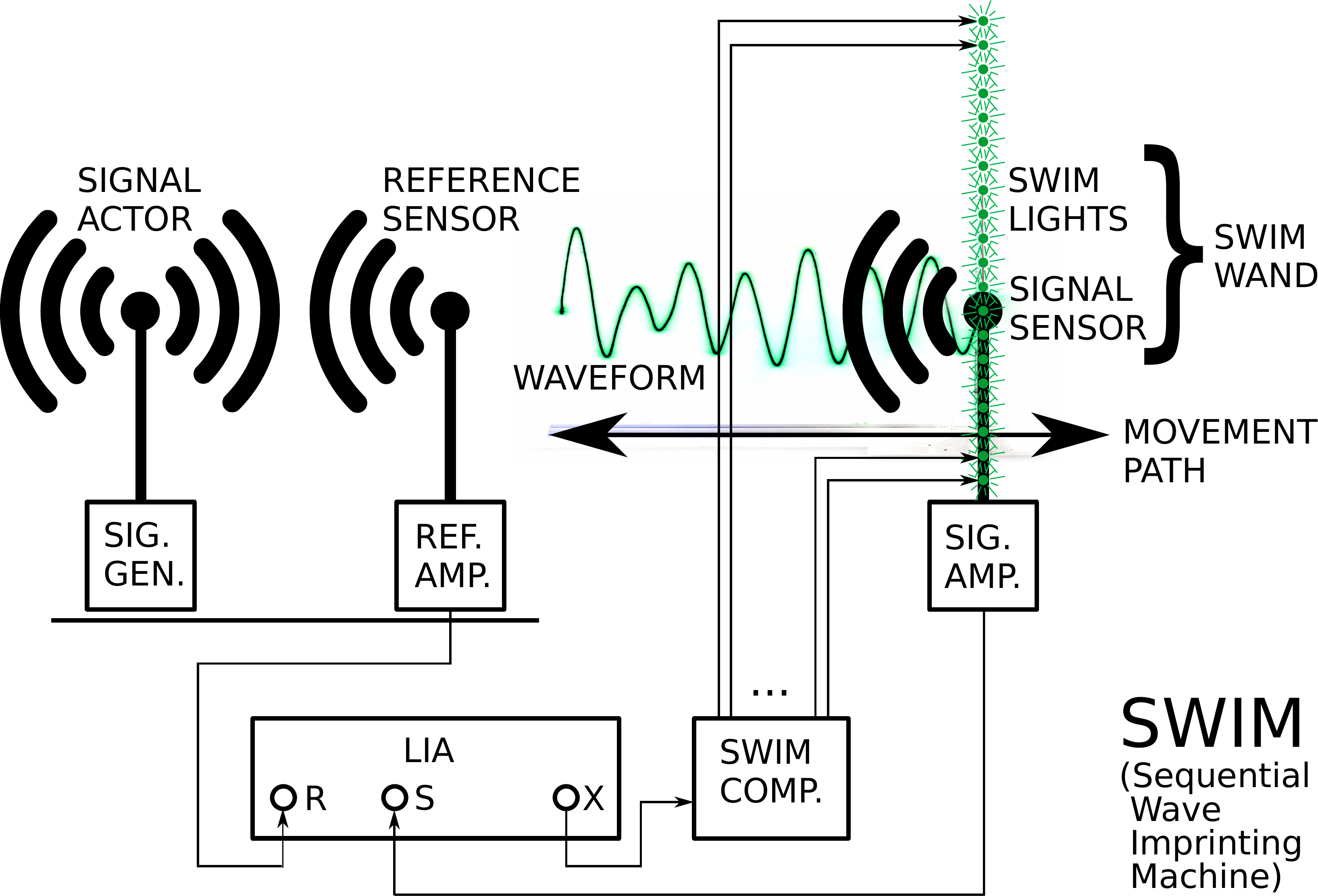}
\hspace{.2in}
\includegraphics[height=2.5in]{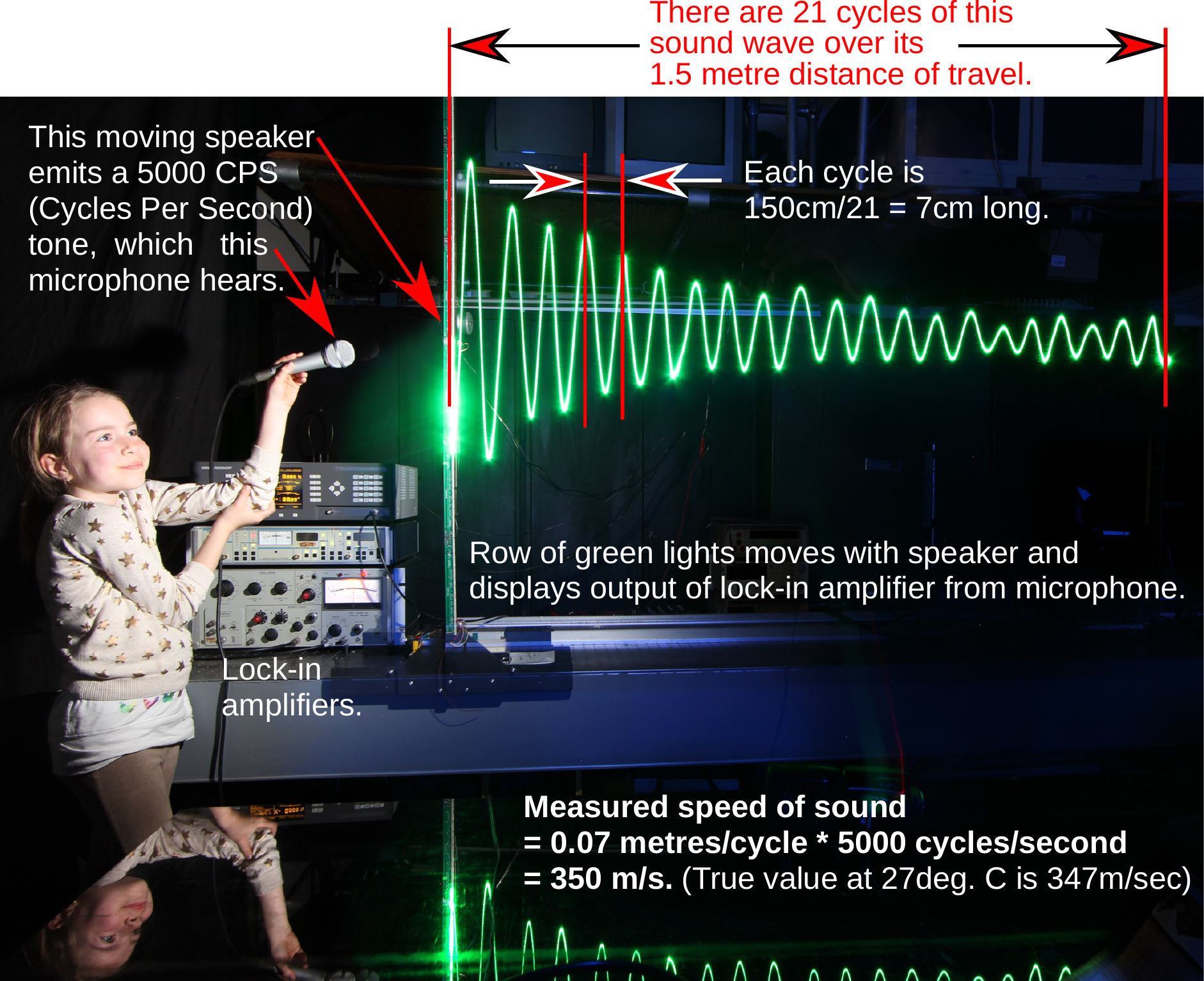}
\caption{SWIM (Sequential Wave Imprinting Machine) principle of operation: A lock-in amplifier (LIA) compares a reference sensor with another sensor that moves together with a linear array of electric light sources.  A long-exposure photograph captures the electric wave (radio wave or sound wave).  Rightmost: Photograph of a sound wave, from which we can measure the speed of sound.}
\label{fig:swimstephanie}
\end{figure*}

Fig~\ref{fig:swimstephanie} shows an example of Phenomenological Reality directed to seeing sound.  Here a linear array of 600 LEDs forms a simple dot-graph indicator of voltage (zero volts at the bottom and 5 volts at the top), and is connected to the output of a lock-in amplifier driven by a moving sound transducer referenced against a stationary sound transducer.  The relative movement between the two transducers causes corresponding relative movement along the phase front of the sound wave, thus making it visible in coordinates in which the speed of sound is exactly zero (Mann 2016).
Because the waveform is driven directly by fundamental physics (e.g. nature itself), we can use it as a scientific teaching tool from which direct measurement may be made.  The array of lights traveled a distance of 1.5 meters, and there are clearly visible 21 cycles of the sound waveform from a 5kHz tone.
Thus each cycle of the 5kHz tone travels 150cm/21 = 7cm in one cycle, i.e. each cycle of the waveform indicates a 7cm wavelength.
From this we can calculate the speed of sound = (0.07 metres/cycle) * (5000 cycles / second) = 350 m/s.

Since the temperature was 27 degrees C, we know from theory that the speed of sound is 347 m/s.  Thus we have an experimental error of approximately 0.86 percent.

This multimediated reality system is useful for teaching physics to children, as well as for making physics, and in particular wave propagation, directly visible. 

\begin{figure*}
\includegraphics[height=.32\textwidth]{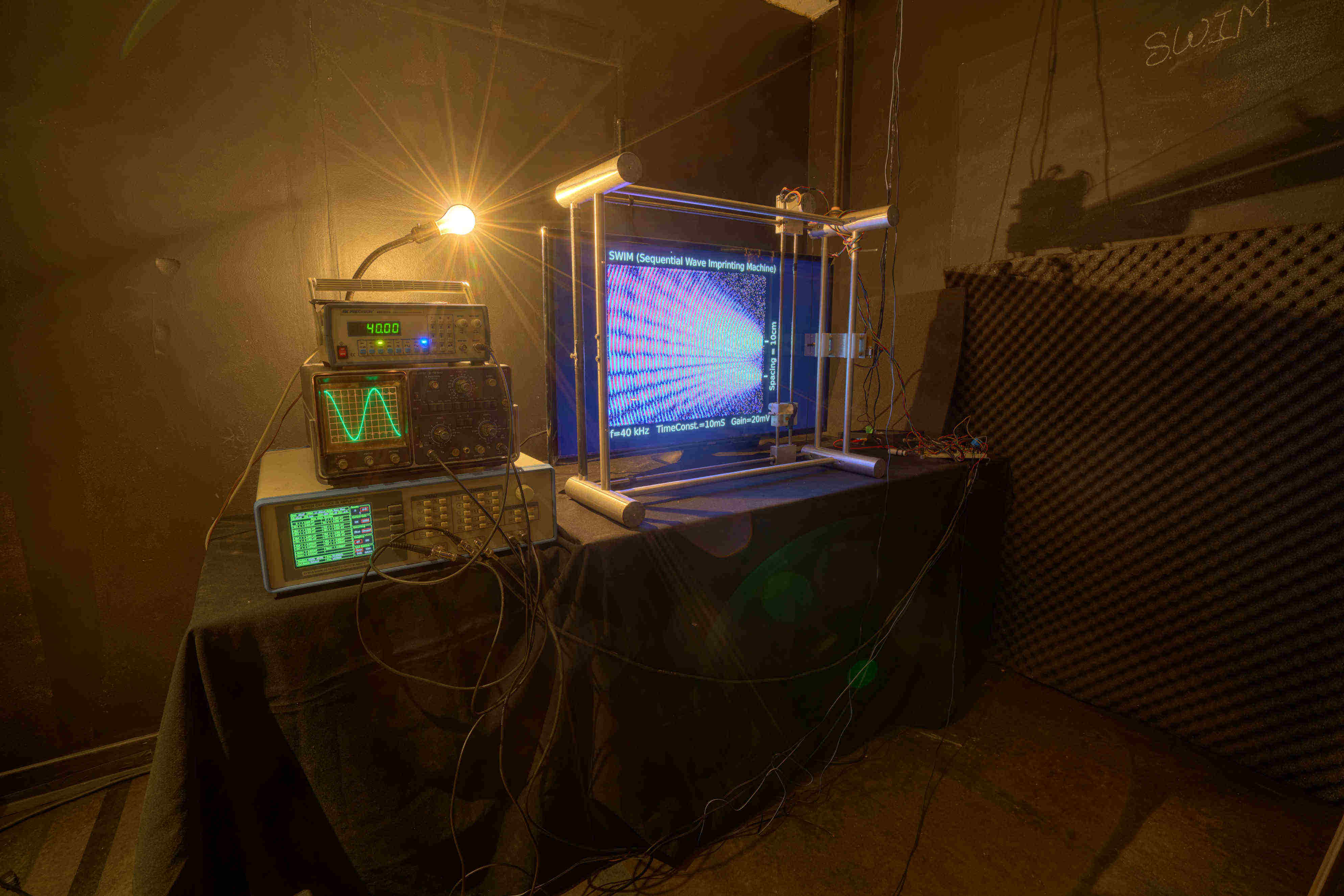}
\includegraphics[height=.32\textwidth]{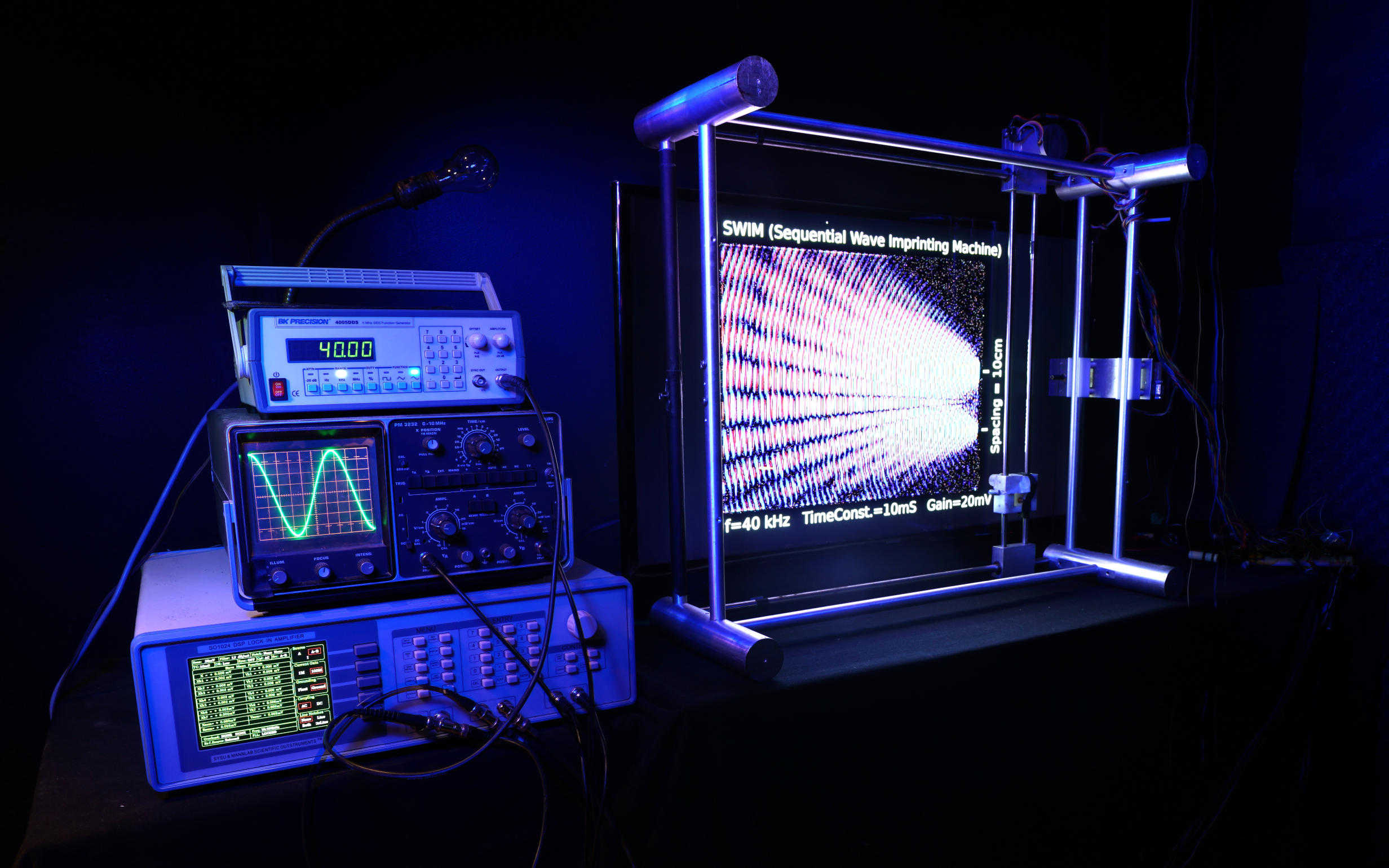}\\
\vspace{.05in}
\includegraphics[width=\textwidth]{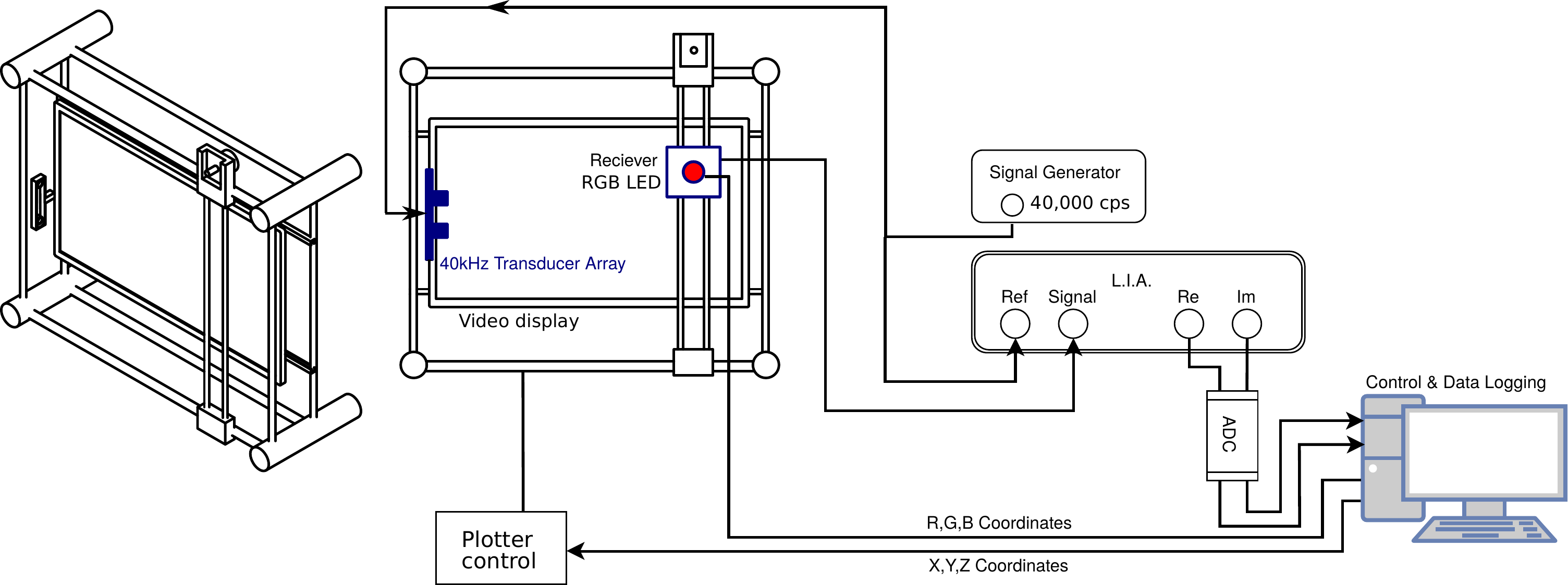}
\caption{Multimediated Reality darkroom with 3D mechanical position control device that scans a space with transducers connected to a special lock-in amplifier (visible in lower left portion of upper right picture).   The signal generator drives the transmit transducer here at 40kHz (signal generator waveform appears on the oscilloscope that's sitting on top of the amplifier).
Top left: with the lights on (HDR photograph); Top right: with the lights off, the sensory attenuation experience helps us concentrate on the multimediated reality that shows interference patterns between two sound sources.
Bottom: Experimental apparatus for multimediated reality.  An XY(Z) plotter carries a listening device (transducer) together with an RGB (Red Green Blue) LED (Light Emitting Diode) through all possible positions in space.  At each position the sound is sensed phase-coherently by way of a L.I.A. (Lock In Amplifier), while sound is produced by a transmit array comprised of two transmitters, receiving the same signal to which the L.I.A. is referenced.  The outputs Re (Real) and Im (Imaginary) of the L.I.A. are converted to RGB values for display on the LED.  A picture is taken of this movement and presented to a video display, to provide a persistence-of-exposure.  Alternatively the video display may be driven directly by data stored in the Control \& Data Logging system.  In this case, it can be animated by multiplication by a complex number of unit modulus, so that the waves on the screen slowly ``crawl'' at any desired speed-of-sound (e.g. the speed of sound can be set to zero or to some small value so as to be able to see it clearly).
}
\label{fig:darkroom}
\end{figure*}

\begin{figure*}
\includegraphics[height=.476\textwidth]{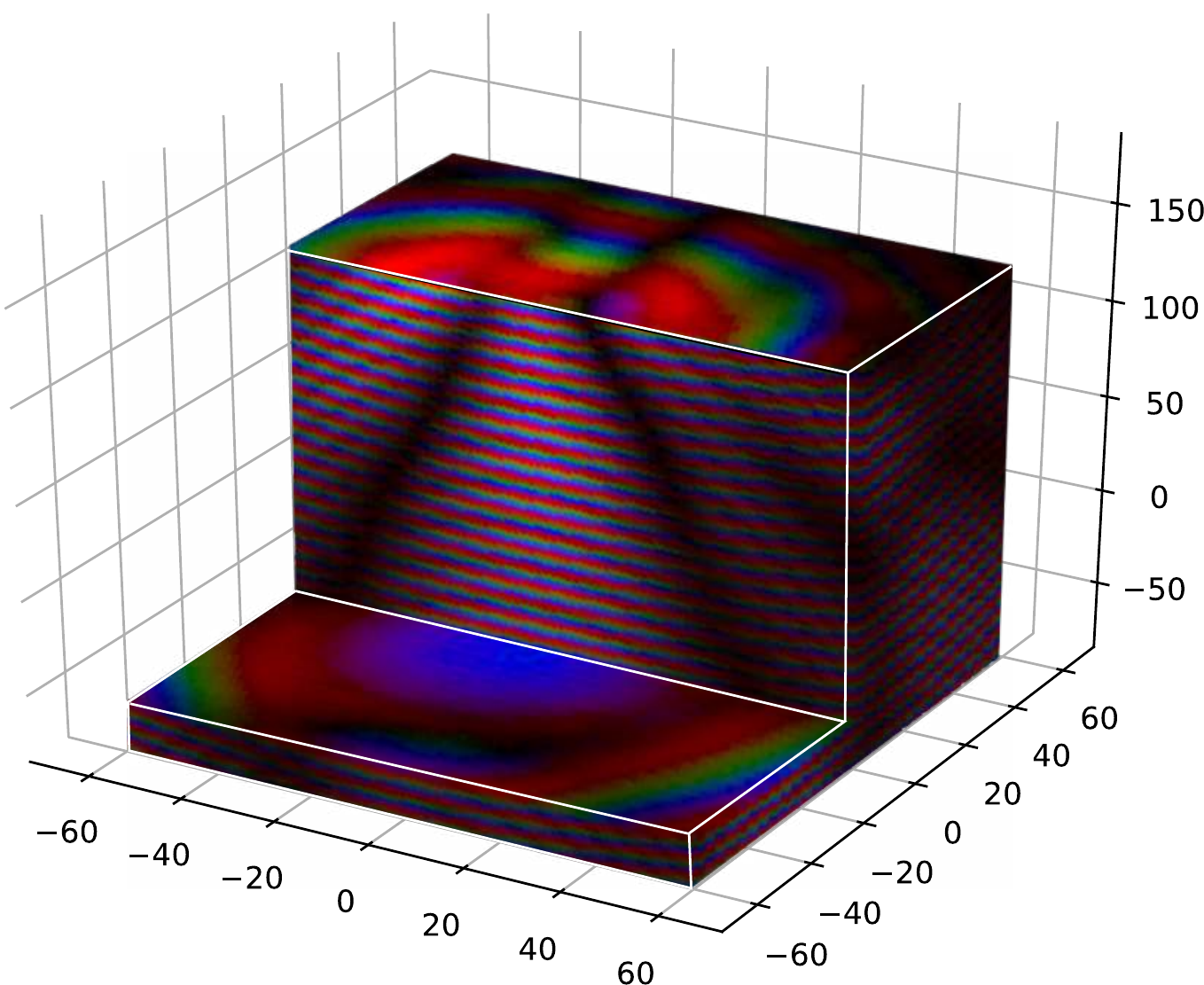}
\includegraphics[height=.476\textwidth]{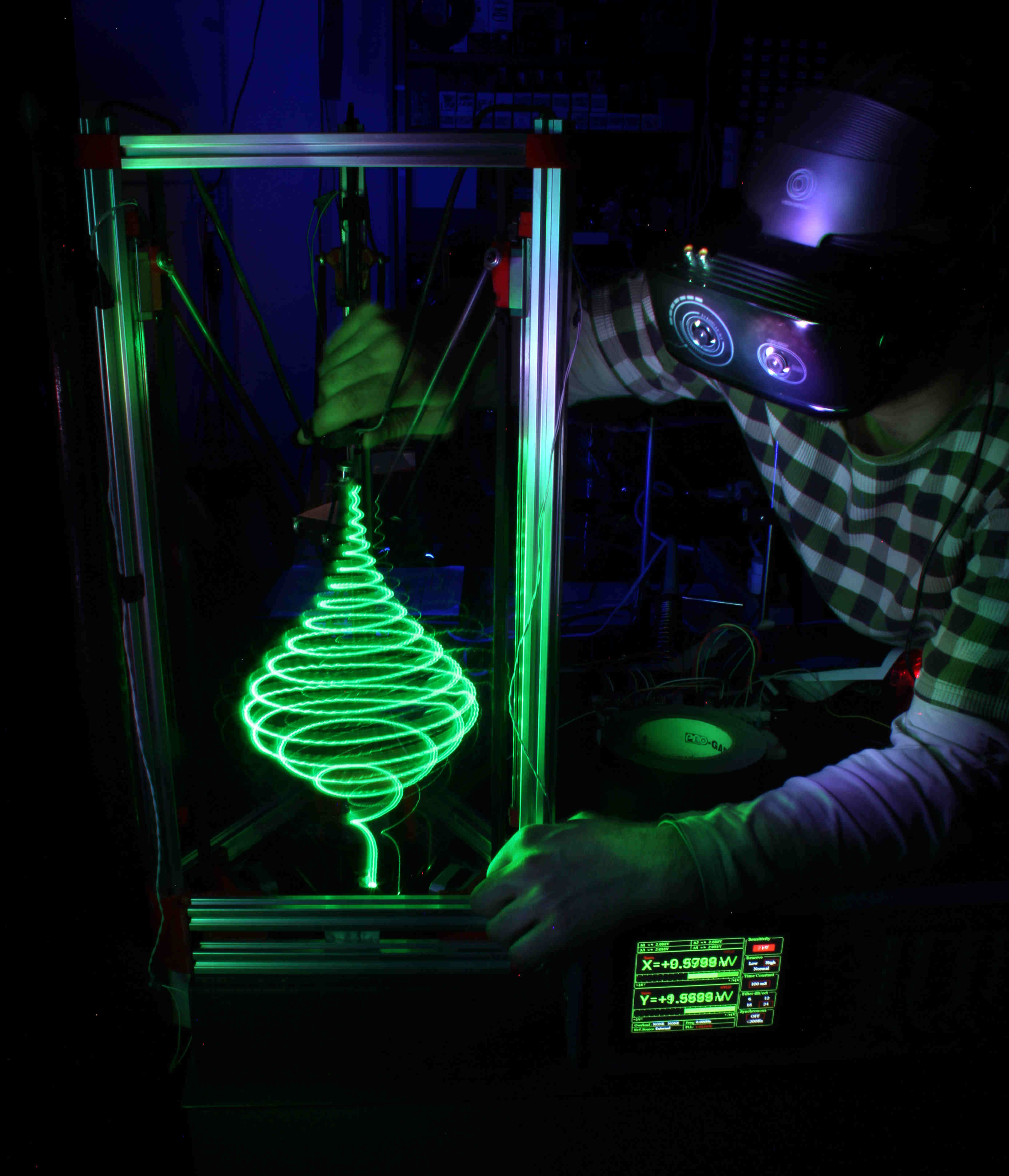}\\
\includegraphics[width=.85\textwidth]{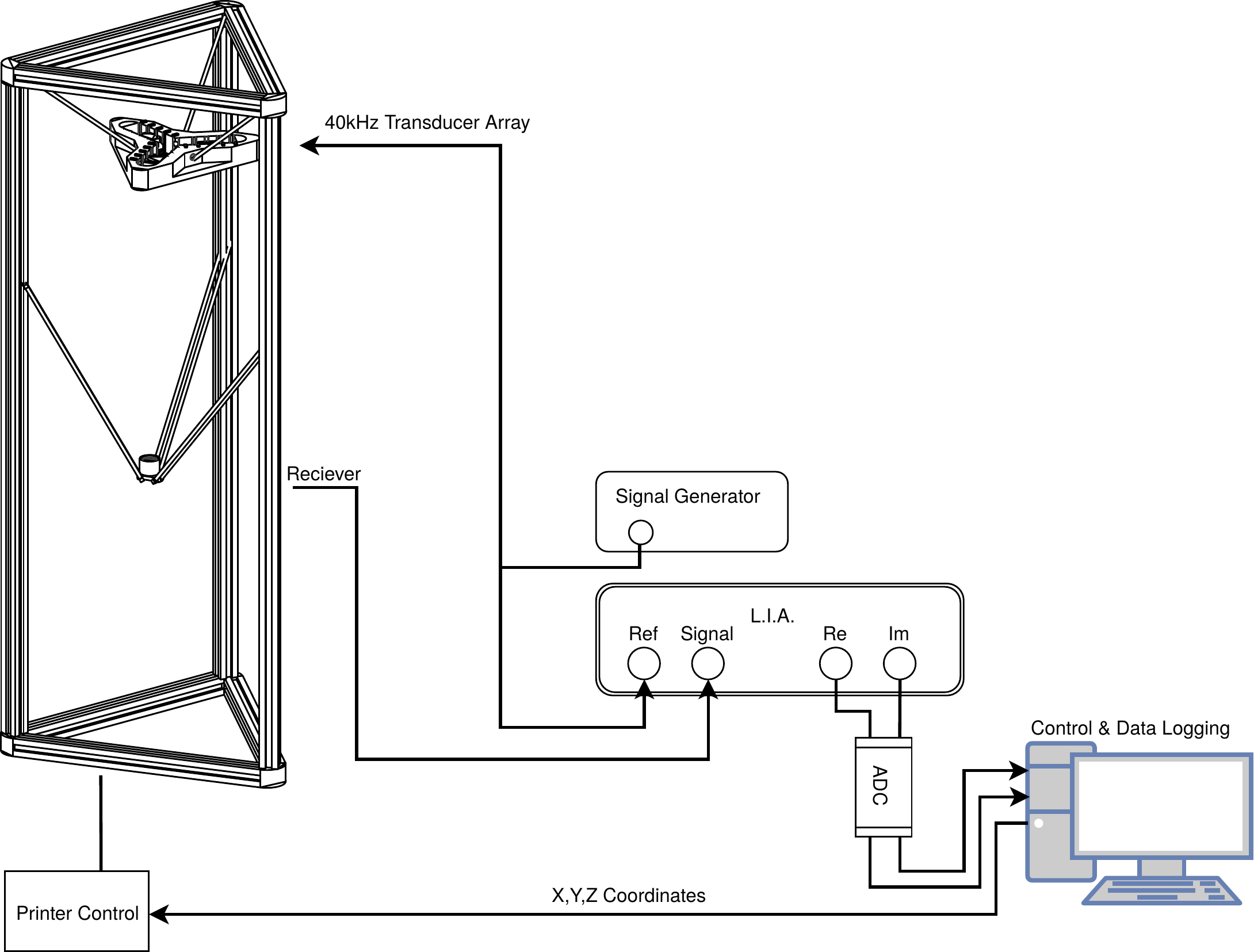}
\caption{With our multimediated reality headset we see the multidimensional interference pattern embedded in 3D space.  Here the interference pattern between three transducers (a triangular array) is shown, and we can also interact with the waveforms by way of patterns traced by the moving print head.}
\label{fig:deltyburn}
\end{figure*}

Multimedia display technologies such as video, as well as special eyeglasses,
can be used to sample and hold the data captured by a moving sensor.
A multimedia darkroom setup of this type is shown in
Fig.~\ref{fig:darkroom} and Fig.~\ref{fig:deltyburn}.
In this way a volumetric dataset is retained as the 3D waveform interference pattern is scanned through space.
While prior SWIM (Sequantial Wave Imprinting Machine) techniques (e.g. long exposure photography) are able to capture 2D and sometimes 3D waveforms, they are recorded from a single two-dimensional perspective. By reading and storing radio wave or sound wave information in point-cloud form, it can be reconstructed, manipulated, and analyzed a-posteriori in three dimensions, allowing for infinite angles and slicing.  This gives a new level of richness to the observation of the wave, and this richness can be explored using the multimediated reality eyeglasses.

\subsection{Comparison with existing measuring instruments}
Many measuring instruments allow us to use one of our senses to measure a quantity that pertains to another sense.  For example, we can have a photographic darkroom light meter that emits sound, so that we can use it in total darkness.  An oscilloscope allows us to see sound waves.  A multimeter is a device that allows us to see or hear electrical signals which would otherwise be difficult or impractical to sense directly.  What is unique about multimediated reality as compared with these traditional measurement instruments is that multimediated reality provides a direct physical alignment between the measured quantities and the real world from which they come.

Multimediated reality allows sound waves or radio waves to be seen in perfect alignment with the world in which they exist, i.e. at 1:1 scale, and ``frozen'' in time (i.e. visible in a set of coordinates in which the speed of sound or speed of light is zero, or a small number that allows the wave propagation to be studied easily).

\section{Multimediated Reality is Multiscale, Multimodal, Multisensory, Multiveillant, and Multidimensional}
\begin{figure}
\includegraphics[width=\columnwidth]{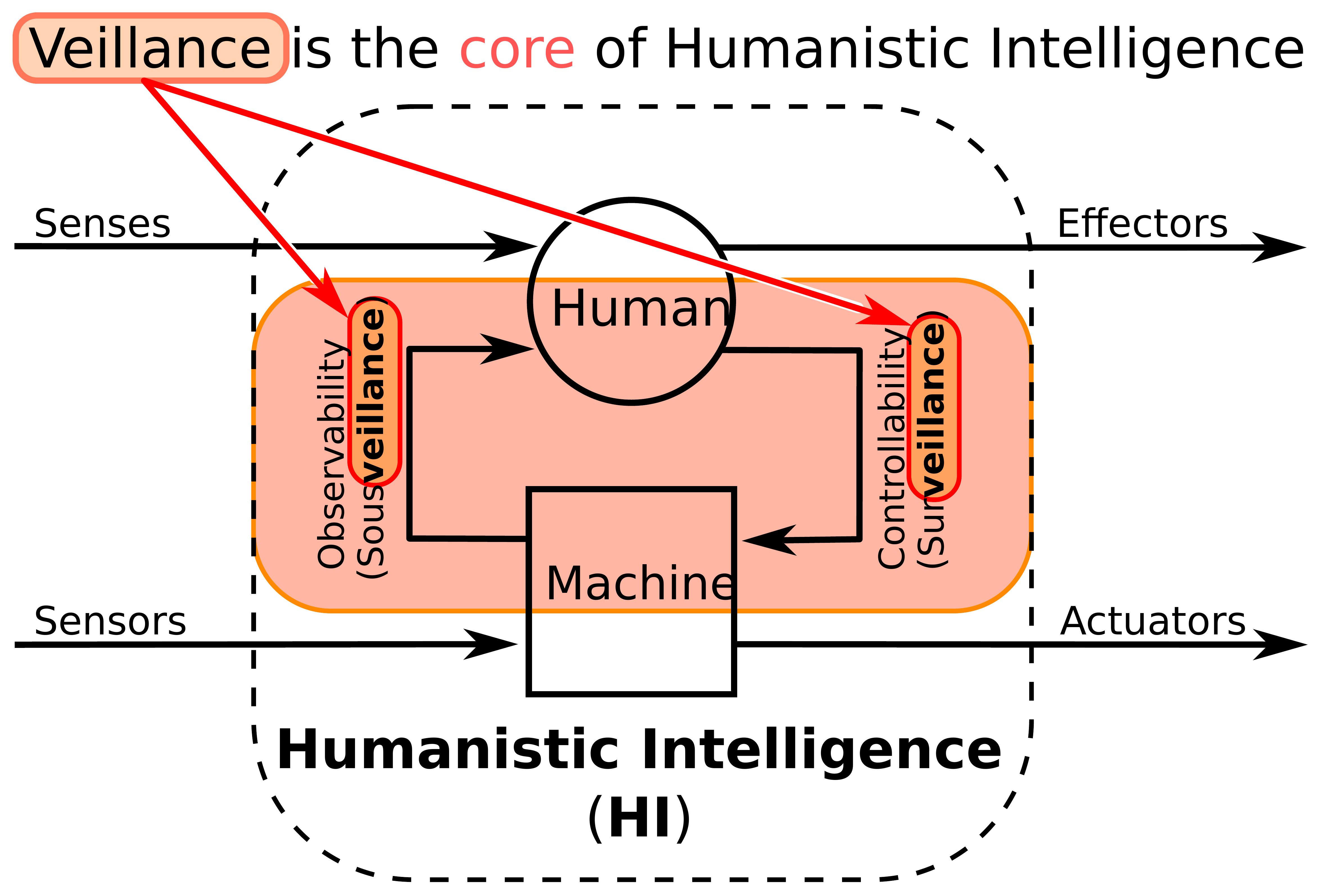}
\caption{An important aspect of multimediated reality is HI (Humanistic Intelligence).  HI addresses systems in which our human senses (sight, hearing, etc.) and effectors (e.g. hands) are augmented or mediated by machine intelligence having sensors and actuators.
HI is intelligence that arises by having the human in the feedback loop of a computational process.  It requires sousveillance (undersight) whenever surveillance (oversight) is present, i.e. it requires a full closed-loop such that if a machine can sense us, we must also be able to sense the machine\protect~\cite{minsky2013society}.  This reciprocity is the core feature of HI that enables it to form a fundamental basis for multimediated reality.
Thus multimediated reality is multiveillant (in contrast to monoveillant technologies that include only surveillance without sousveillance).
}
\label{fig:hi}
\end{figure}

Multimediated reality is more than just a taxonomy of real and synthetic experience.  It also considers how we interact with the world around us and each other, through the use of technology as a true extension of our own minds and bodies.  Specifically we consider the concept of AI (Artificial Intelligence) as well as
human-in-the-loop-AI, also known as HI (Humanistic Intelligence)~\cite{minsky2013society}.
HI posits that technology should function as an intermediary between us and our environment in such a way that the intelligence it affords us arises through a computational feedback loop of which we are a part.  See Fig.~\ref{fig:hi}

\begin{figure*}
\includegraphics[width=\textwidth]{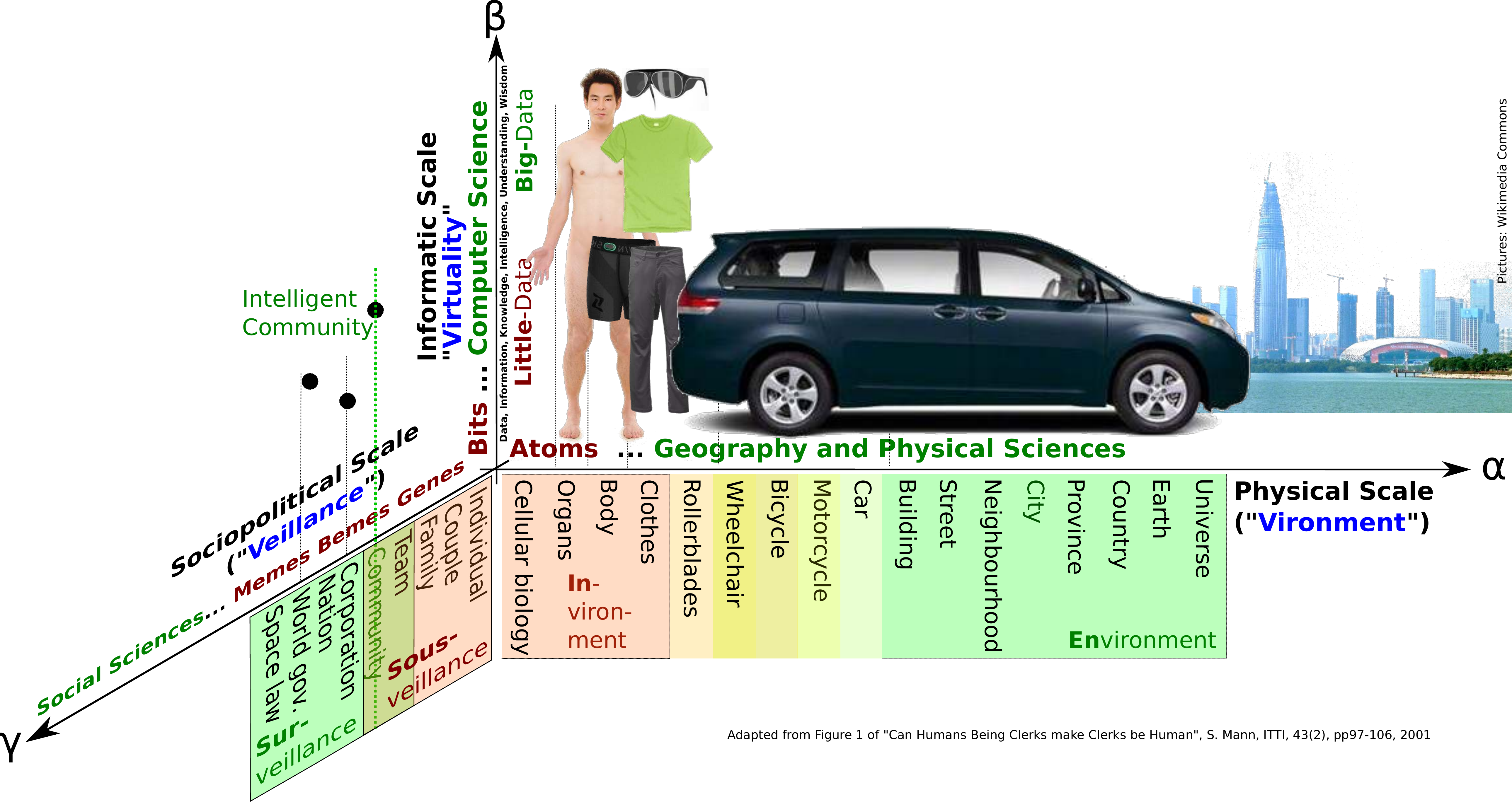}
\caption{Multimediated reality is multiveillant (surveillance AND sousveillance) as well as multiscale (wearables AND smart environments).
We can identify at least three axes: Firstly, a physical scale axis (of physical reality) defines the environment (that which surrounds us) and the invironment (us ourselves).  At the border between the environment and invironment are things like clothes, and other technological prostheses.
A Virtuality axis defines also a scale from ``Bits'' all the way out to ``Big Data''.
A sociopolitical or ``Veillance'' axis defines sousveillance (individual/internal) out to surveillance (external).
At the origin are ``Bits'', ``Atoms'', and ``Genes''.  Genes, for example, are the smallest unit of ``humanness'' (human expression).
The axes are labeled $\alpha$, $\beta$, and $\gamma$.
The first of the three axes ($\alpha$) is denoted pictorially, at physical scales starting from naked, to underwear, to outerwear, to a vehicle (car), to the ``smart city''.
}
\label{fig:urbanscalespace}
\end{figure*}

Multimediated reality involves multiple physical scales, including both wearable technology as well as technology in the environment around us, like smart rooms (e.g. smart darkrooms).
This multiscale and multiveillant nature of multimediated reality is illustrated in Fig.~\ref{fig:urbanscalespace}

\subsection{Multisensory Synthetic Synesthesia}
Synesthesia is a neurological condition
in which there is crosstalk between human senses,
e.g. chromesthesia which is hearing colors of light, or
"The Man Who Tasted Shapes"~\cite{Cytowic93}.

Multimediated reality often involves a multimedia-induced (synthetic) synesthesia among and across our existing senses (e.g. seeing sound), or, extrasensory, i.e. beyond our existing senses (e.g. seeing or feeling radio waves).
In this way, multimediated reality is multisensory and multimodal.

\subsection{Multidimensional Multimediated Reality}
Synthestic synesthesia of extra senses like electromagnetic radio waves and radar (or sonar) provides us with a richly multidimensional perceptual space, in the context of multimediated reality.  Whereas existing virtual reality systems might use radio waves or sound waves for tracking and position sensing, they do not directly engage the waveforms
phase-coherently as part of the resulting output space.
When we engage with the radio waveforms or sound waveforms directly, we have many new sensory possibilities for direct experience of multidimensional signal properties like multi-polarization and multipath propagation.

Even simple radio waves become complex-valued when they are brought down to baseband (e.g. when experienced in coordinates where the speed of sound or light is zero).  In this case, the chosen synesthesia is to be able to see complex numbers on a color wheel where phase is color and magnitude is the quantity (photoquantity~\cite{mann2002painting}) of light presented~\cite{rattletale}.

\begin{figure*}
\includegraphics[width=0.49\textwidth]{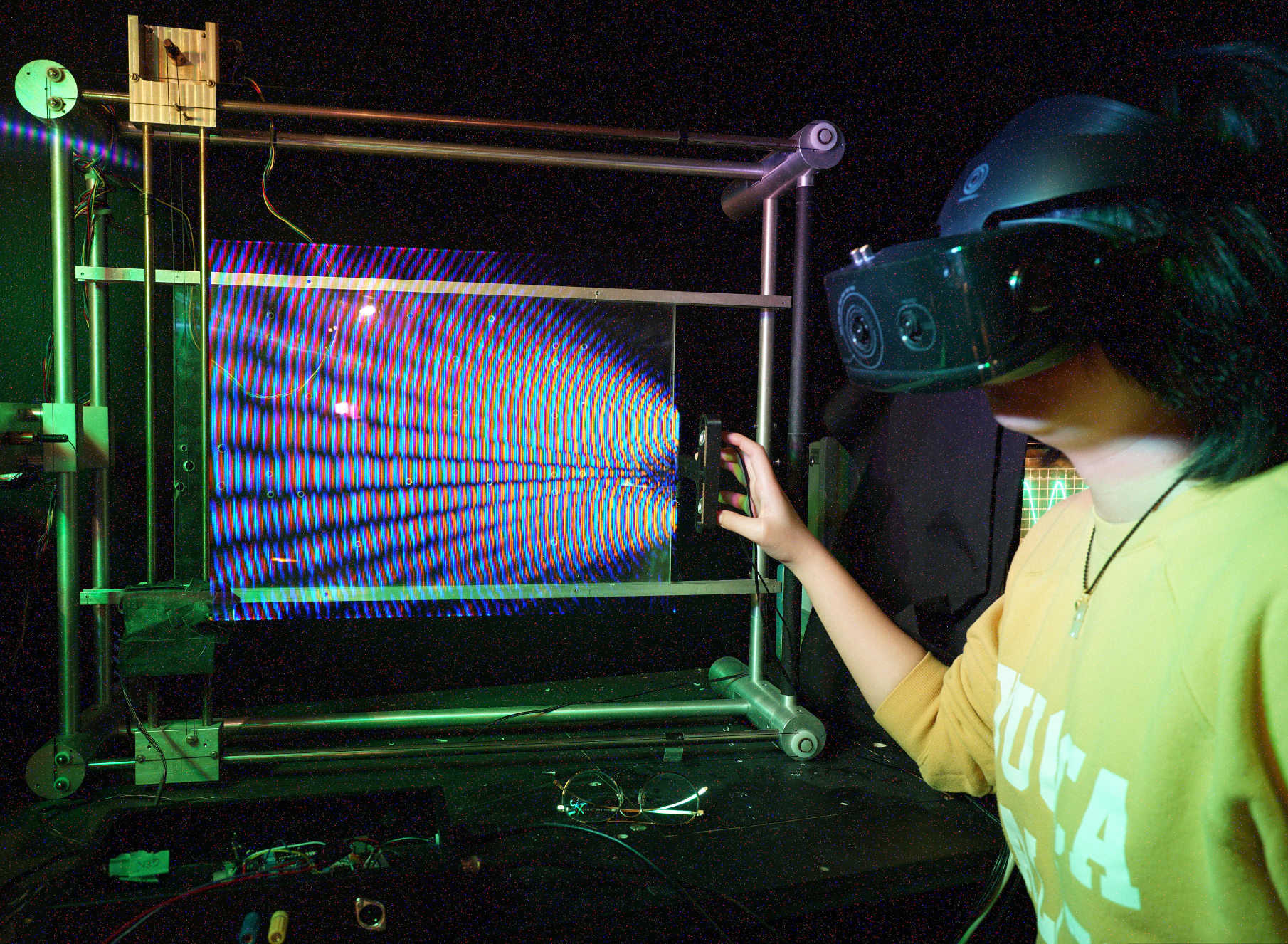}
\hspace{.006\textwidth}
\includegraphics[width=0.49\textwidth]{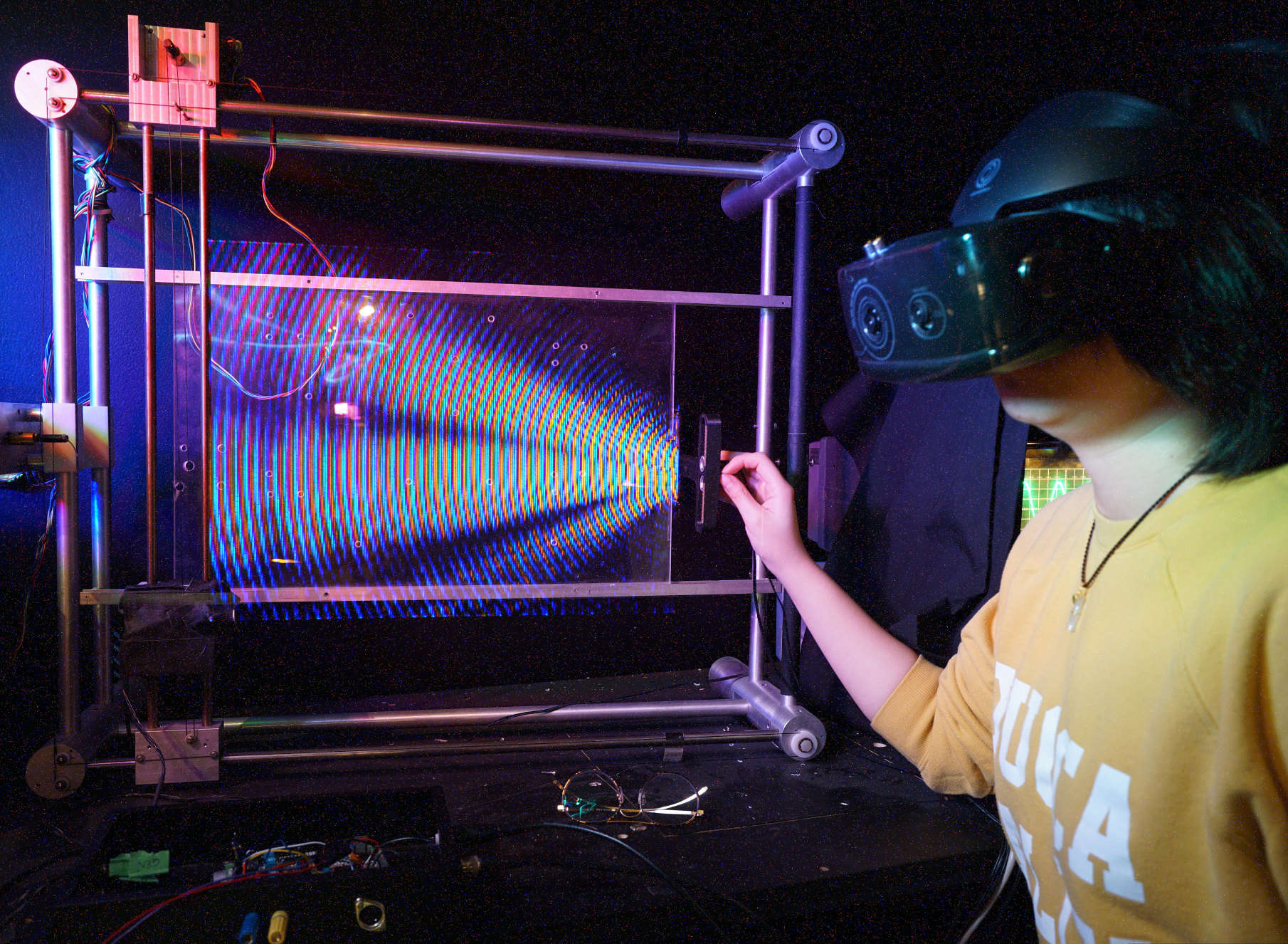}
\caption{Interference pattern of two acoustic wavefronts made visible using the multimediated reality darkroom of Fig.~\protect\ref{fig:darkroom}.
As the two sound sources (speakers) are moved closer together, their acoustic interference fringes spread further apart, confirming a simple law of physics.
Left photograph: wide spacing between speakers gives closely spaced interference pattern.
Right photograph: narrow spacing between speakers gives broad spaced interference pattern.
Sound waves from two movable speakers emitting the same pure tone are photographed with a lock-in amplifier driving  an RGB (Red, Green, Blue) light source moving in space (X, Y, Z). This is not computer graphics!  These are photographs of an actual physical process generated by nature itself, and is merely facilitated (made visible) by computation.
Those wearing a multimediated reality headset see patterns in a high-dimensional space, whereas those without headsets can still see the sound waves, but at a reduced-dimensionality.}
\label{fig:interference2}
\end{figure*}

We see this in Fig.~\ref{fig:interference2}, where the phase of the sound waves is displayed as color.
Thus we can see clearly the interference patterns of two sound waves where they interfere constructively and destructively, as variations in the quantity of light, and the phase fronts, as variations in color.

\begin{figure*}
\includegraphics[width=.75\textwidth]{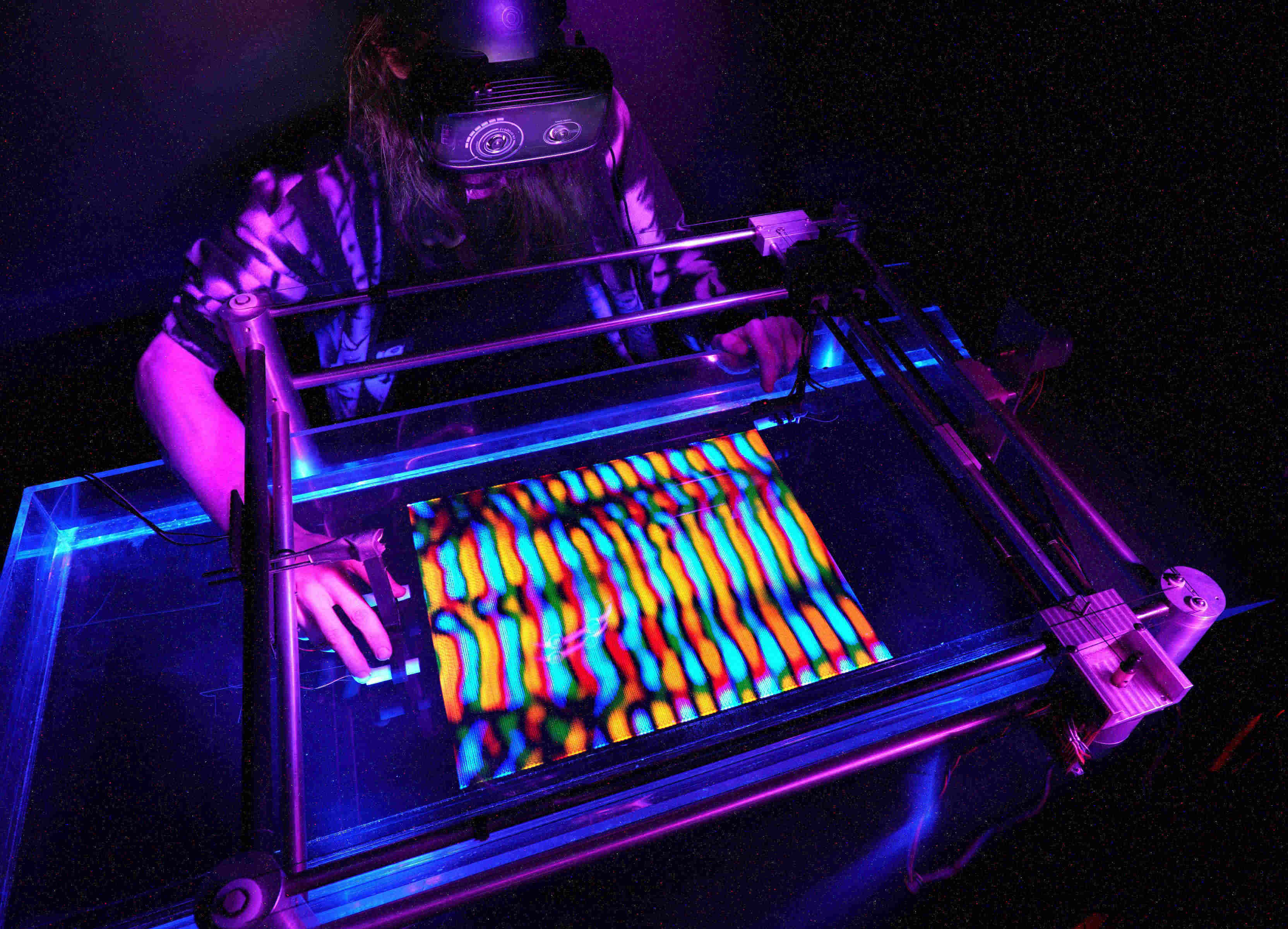}
\caption{Photograph of apparatus showing underwater sound at 40 kHz from two hydrophones, toward the left (the spacing is being controlled by the user's right hand), and a test/reference hydrophone toward the right (near the user's left hand).
The wearer of the multimediated reality eyeglass sees in a high dimensional space (3D spatially, plus time animation, plus complex-valued quantities, etc.).
Others see only a flat 2D slice through the interference pattern.
}
\label{fig:underwaterswim}
\end{figure*}

This system may also be used to study underwater sound waves.  Fig.~\ref{fig:underwaterswim} shows a sonar array being explored.
We have also constructed a Multimediated Reality aquatics facility for use with underwater MR eyewear.  In water, the weightless state allows us to feel as if we are floating among the underwater sound waves.

\section{Multimediated Reality Continuum}
\begin{figure*}
\includegraphics[width=.9\textwidth]{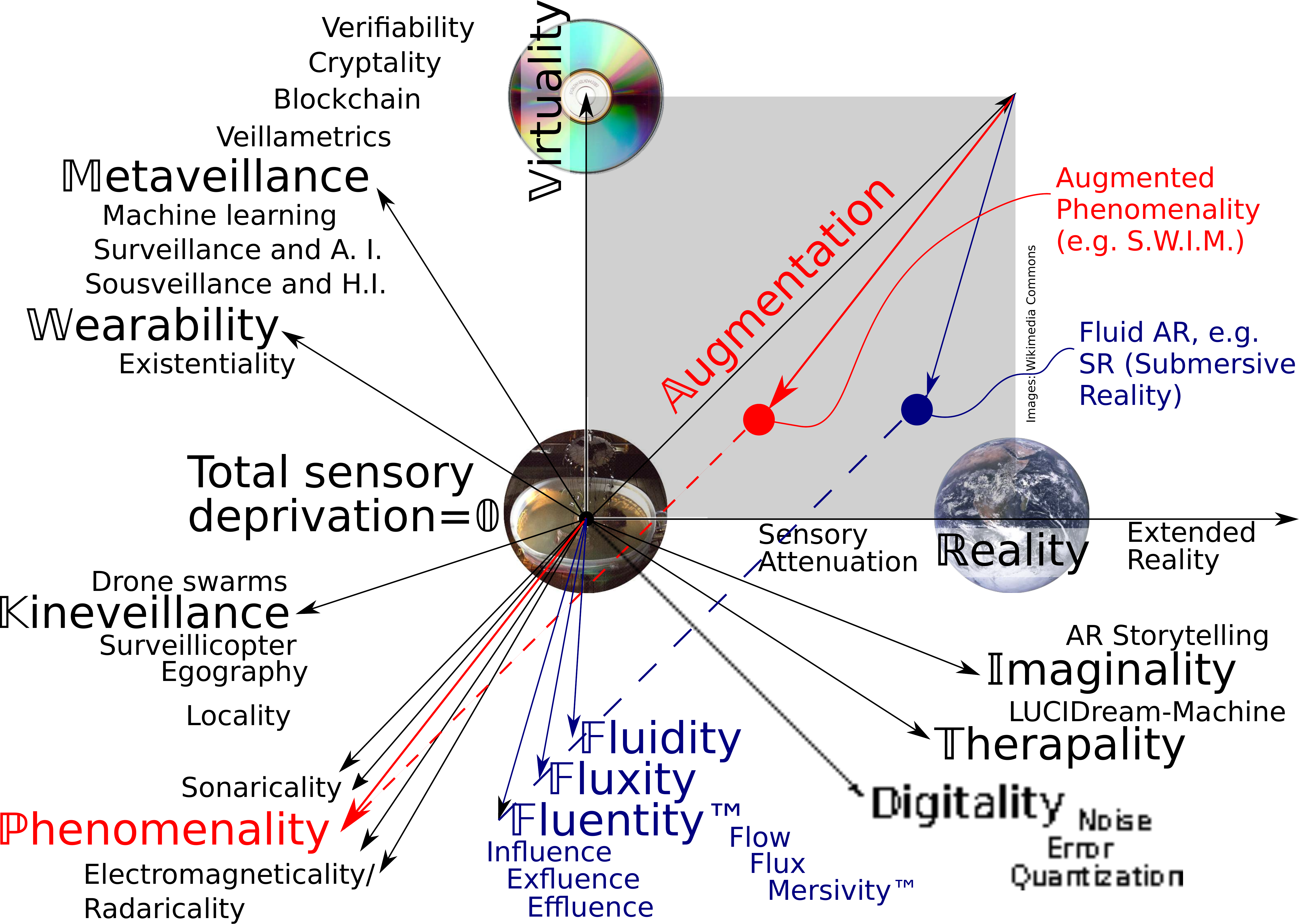}
\caption{The Multimediated Reality Continuum.  Reality is the main axis going from left-to-right, starting at ``Total sensory deprivation'' (the origin, indicated by a sensory deprivation tank), then to ``Sensory attenuation'', then Reality, and then beyond Reality to give also Extended reality.
Virtuality is the secondary axis pointing upwards.  Augmentation exists in the 2-dimensional space spanned by Reality and Virtuality.
A third axis, phenomenality, indicates any kind of phenomenological reality, such as phase-coherent photography of radio waves or sound waves, such as by
Sequential Wave Imprinting Machine (SWIM).  In this sense, PAR (Phenomenological Augmented Reality)~\protect\cite{mann2015par}  is a combination of AR and P (Phenomenality).  A point in this space is indicated by the red dot as ``Augmented Phenomenality''.
As another example, consider a point (indicated in blue) that comes out from AR along the Fluentity axis.  An example of this kind of reality is the Internet-based underwater virtual/augmented reality performance space~\protect\cite{icmc2007performance,teletubs}.
When we submerge ourselves in a large swimming pool, with an underwater SWIM to see sound waves (e.g. to test out an underwater sound system and see the interference patterns between two underwater speakers), we're in a reality described by adding the two vectors (red dot and blue dot) taking us into an additional higher dimension.
This works like a giant version of our apparatus of Fig.~\protect\ref{fig:underwaterswim} in which we submerge ourselves fully in the pool while wearing the All Reality headset.
The All Reality Continuum thus allows us to understand sensory attenuation (interpolation between sensory deprivation and reality) as well as eXtended reality (extrapolation beyond reality), in addition to the many other dimensions shown, such as Metaveillance\protect~\cite{kineveillance} (Sur/Sous-Veillance, smart cities, smart buildings, etc.),
Wearability (Mann's ``Wearable Reality'' of 1974 and Canadian Patent 2388766), Kineveillance (drone and swarm-based realities\protect~\cite{mann2014introduction}), Imaginality (e.g. lucid dreaming in the sensory deprivation tank), and Therapality (the axis of lasting effects that persist even after shutting off and removing a technology).
Not all axes are desirable, e.g. Digitality is the axis of quantization noise that embodies the undesirable artifacts of {\em being digital}, pixelation, etc.  Ideally we wish to use computers for {\em being undigital}~\protect\cite{mannwyckofftr}.
}
\label{fig:multimediatedcontinuum}
\end{figure*}

Many of the systems presented in this paper do not fit nicely into existing
taxonomies of VR and AR, or any of the more general taxonomies of synthetic experience~\cite{robinett1992synthetic}.
We proffer a more general ``reality'' continuum
in which the space is multidimensional,
and in which the origin is the absence of sensory stimulation,
allowing us to consider technologies such as
sleep masks,
interactive sleep masks,
sensory deprivation tanks, interactive sensory deprivation tanks~\cite{teletubs,icmc2007performance}, aquatics facilities,
theatres, darkrooms, therapy systems, 
and the like, as a basis upon which to
create new realities directly connected to physical or intellectual phenomena.
See Fig.~\ref{fig:multimediatedcontinuum}
Note the many dimensions and the many ways they can be combined.  For example we can have a mix of Reality and Virtuality that gives AR (Augmented Reality), and then further add some phenomenality to get PAR (Phenomenological Augmented Reality~\cite{mann2015par}).
We can add to AR some Fluentity to get SR (Submsersive Reality~\cite{teletubs,icmc2007performance}).
And if we do PAR while swimming fully submerged in water, we're spanning the four dimensions of Reality, Virtuality, Phenomenality, and Fluidity/Fluentity.

Note also that many of the dimensions are inherently combined and thus certainly do not form an orthonormal basis.  For example, Metaveillance includes Surveillance (oversight) and AI (Artificial Intelligence), as well as Sousveillance (undersight) and HI (Humanistic Intelligence).
Sousveillance and HI are very closely related to Wearability.  Thus there is strong overlap between Metaveillance and Wearability.  Wearability is closely correlated to Existentiality~\cite{itti}, giving us the ``Wearable Reality'' (Canadian Pat. 2388766) proposed by Mann in 1974 as embodied by the SWIM (Sequential Wave Imprinting Machine), itself an example of Phenomenality~\cite{mann2015par}.  Thus these three axes, Metaveillance, Wearability, and Phenomenality, are closely related.

Not all dimensions are desirable.  For example, sensing can be done in a discrete (quantized) or continuous manner.
We prefer systems in which computers are used to sense ``undigitally'', i.e. above the Nyquist rate spatiotemporally.  Today's digital cameras do this, for the most part, but still exhibit tonal/level quantization, thus the need for {\em being undigital~\cite{mann1994beingundigital}} with digital cameras, e.g. HDR sensing and metasensing~\cite{scourboutakos2017phenomenologically,mann2015par}.
Ideally we would like to use computers for {\em undigital senses / sensors} as well as for {\em undigital effectors / actuators}, i.e. for all six signal flow paths of Fig.~\ref{fig:hi}.
This allows us to achieve undigital experiences, i.e. what we proffer to call an {\em undigital reality}, which is a multimediated reality that is free from the artefacts of digital computation.  This is especially important for SR (Submersive Reality) in which the free flowing nature of swimming makes it possible to forget that one is wearing a VR/AR/MR headset and imagine, in the totally weightless world, another reality that is totally fluid and free of ``{\em digitalness}'' (quantization noise, pixelation, etc.), a world where digital senses / sensors,
or the like would be totally unacceptable.

\subsection{Multimediated Reality is ``\raisebox{-.38ex}{{\tt {\Huge *}}}R'' (All R)}
In most search engines and computing environments, the asterisk symbol, ``\raisebox{-.175ex}{{\tt {\LARGE *}}}'', is a ``wildcard'' that can mean anything.  It can be replaced with any other characters, symbols, words or phrases, and dates back as far as the TOPS-10 operating system in 1967.  Thus for the space defined by the Multimediated Reality continuum, we proffer to call it ``\raisebox{-.25ex}{{\tt {\LARGE *}}}R'', pronounced ``All R'' (all realities).  In situations where there is only room for two characters, and where an asterisk cannot be used conveniently (e.g. file names in the Unix operating system), we proffer ``ZR'', to emphasize the possibility of complex-valued multidimensional data of the form $Z = X + i Y$ where $i=\sqrt{-1}$ is the imaginary unit.
We must realize, though, that there are many dimensions, and some of these many dimensions may be complex-valued, or beyond, i.e. not limited to the two-dimensions implied by the Argand plane
(https://en.wikipedia.org/wiki/Complex\_plane).

\section{Summary and Conclusions}
We have presented Multimediated Reality as a proper superset of mediated (XY), mixed (X), augmented (A), and virtual (V) reality (R).

Multimediated reality uses interactive multimedia in a way that is:
\begin{itemize}
\item Multidimensional, in a true interactive sense, i.e. direct interaction with concepts like cross-polarization, and complex-valued multidimensional mathematical quantities, e.g. Fieldary User Interfaces [Gerson et al 2015];
\item Multisensory, cross-sensory (e.g. synthestic synesthesia);
\item Multimodal, in many senses of the word, e.g. multimodal interaction (human-computer interaction using multiple modalities of input/output), and
multimodal artifacts (artifacts that use multiple media modes, including
social media, gaming, storytelling, etc) [Gunther Kress, 2010. Multimodality: A Social Semiotic Approach to Contemporary Communication. New York: Routledge];
\item Multidisciplinary (i.e. as a field of inquiry, involving the arts, the sciences, engineering, medicine, health, wellness (e.g. meditation in multimediated sensory attenuation tanks), etc.
\end{itemize}

\begin{figure}
\includegraphics[width=.6\columnwidth]{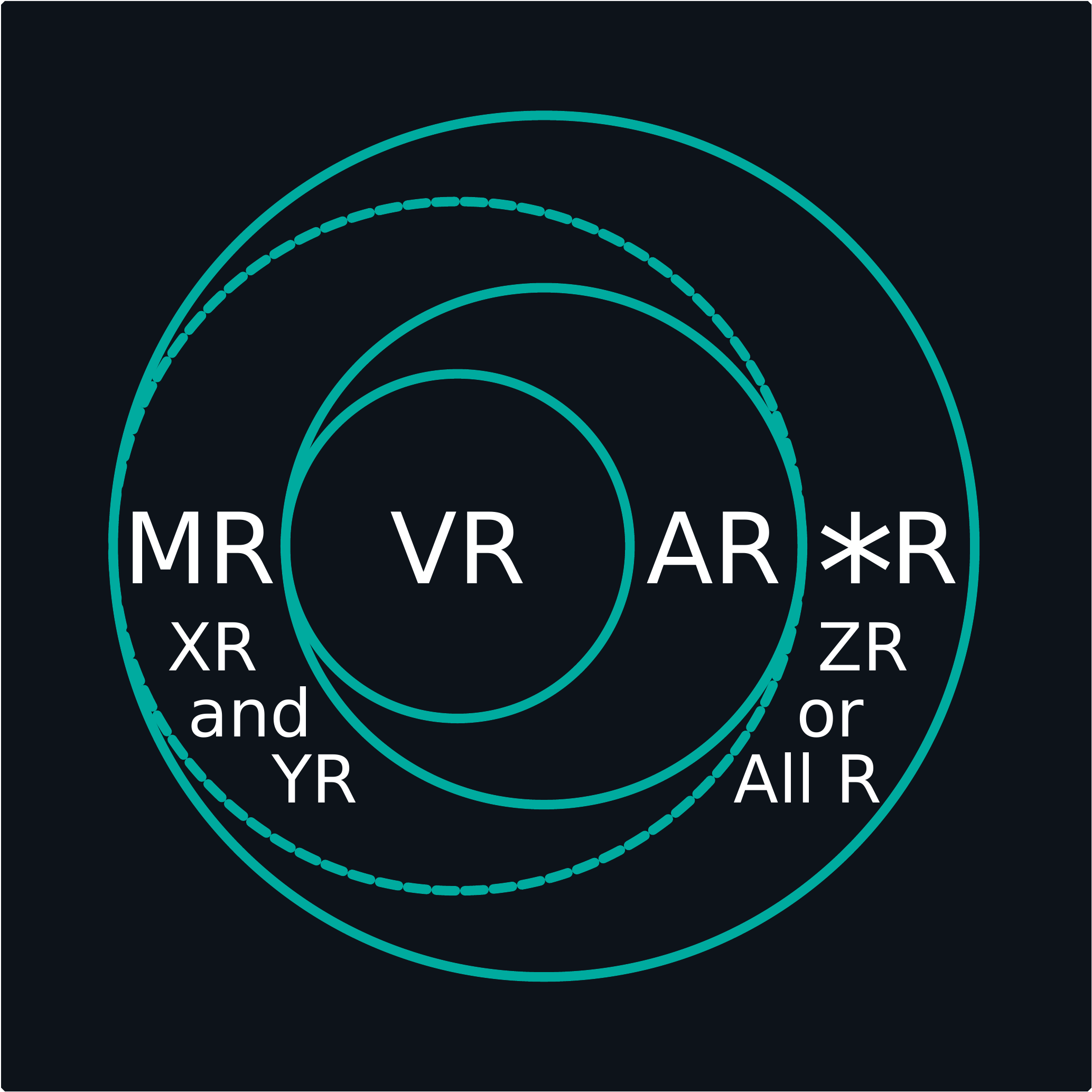}
\caption{Venn diagram showing various realities.  VR is a proper subset of AR.
AR is a proper subset of mixed reality.  Mixed reality and X-Reality\texttrademark are proper subsets of mediated reality (mixed, X-, and mediated/Y- reality are denoted by a dotted line in the diagram in order to indicate that either could be represented by this circle).  These realities are a proper subset of multimediated reality, which we
proffer to call ``\raisebox{-.175ex}{{\tt {\LARGE *}}}R'', pronounced ``All R'', where ``\raisebox{-.15ex}{{\tt {\LARGE *}}}'' denotes any of the other forms of iteractive multimedia reality.
Alternatively, e.g. in systems like the UNIX operating system where
``\raisebox{-.15ex}{{\tt {\LARGE *}}}'' is a special reserved character, we can abbreviate ``All R'' as
``ZR'' to emphasize its multidimensional and complex-valued nature, i.e.
Z = X + iY, where i is the imaginary unit number, $i=\sqrt{-1}$.
%Right: Our Multimediated Reality eyewear prototype that senses and displays multidimensional reality.
}
\label{fig:combined}
\end{figure}

\begin{figure}
\includegraphics[width=\columnwidth]{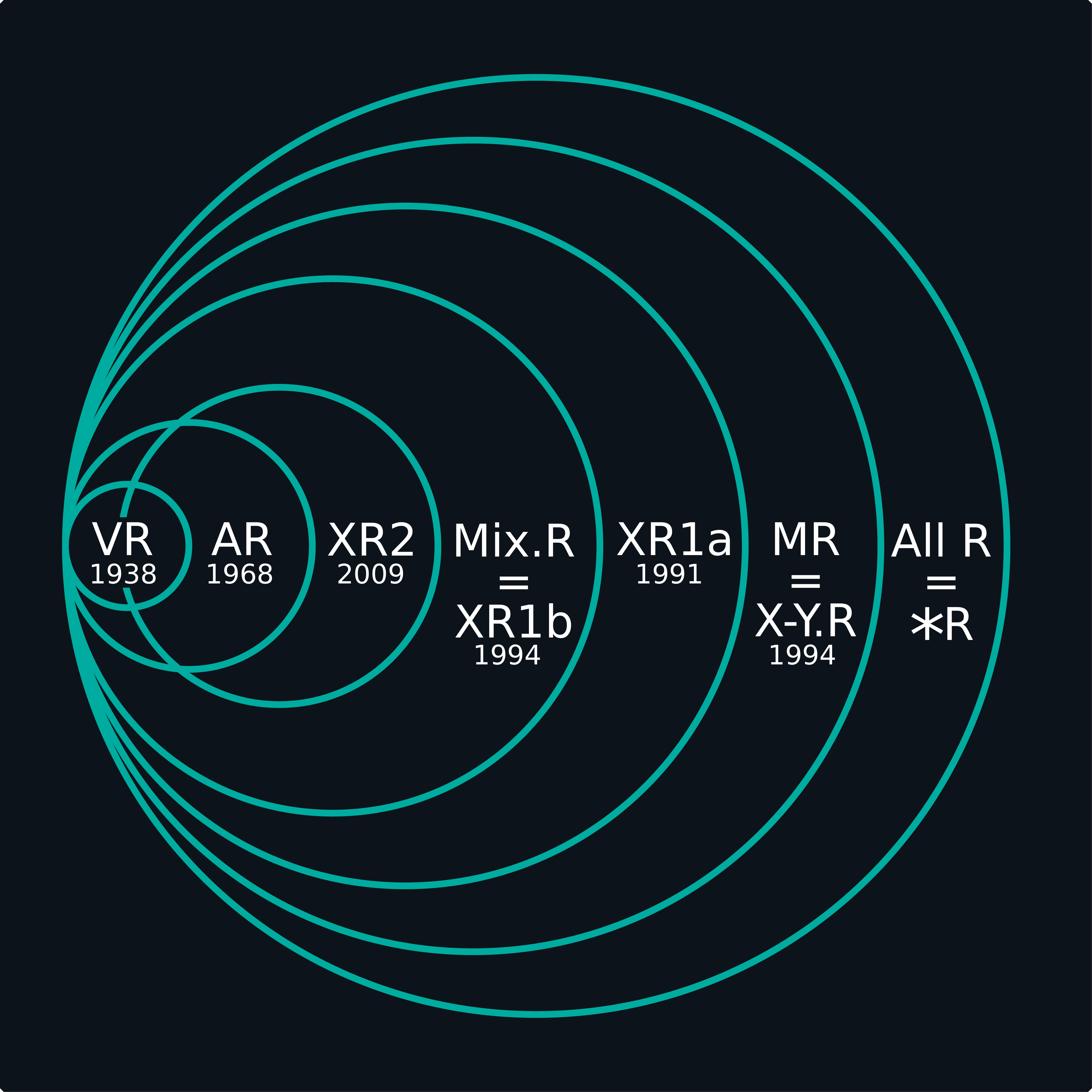}
\caption{A taxonomy of seven realities, together with the approximate year each was first introduced, bearing in mind that some of them evolved over many years (e.g. consider that VR was introduced by Artaud in 1938, AR was introduced by Sutherland with his see-through HMD in 1968, etc.).}
\label{fig:VR_AR_MR_XR_ZR}
\end{figure}

Combining Fig.~\ref{fig:xr} and Fig.~\ref{fig:combined},
we have mostly a nested set except for XR2 which is a proper subset of mixed reality, i.e. limited to a specific kind of virtual world.
See Fig.~\ref{fig:VR_AR_MR_XR_ZR}.

Additionally, we have presented some new forms of multisensory multidimensional interactive reality, such as multidimensional complex-valued fields and other
quantities that are captured and displayed to the human senses or recorded on
photographic media, video display media, or wearable media.  Whereas virtual reality and other realities (e.g. augmented reality) usually show artificial graphical content (computer graphics), multimediated reality also has the capacity to show a ``real reality'', i.e. to make visible what is really present (but otherwise invisible) in the world around us.

\begin{acks}
Authors would like to acknowledge the many students who were involved with the project: Max, Sen, Jackson, and former students Arkin and Alex, as well as Kyle.
The work is supported by Mannlab Shenzhen.
\end{acks}